\documentclass [prd,nofootinbib,eqsecnum]  {revtex4}
\usepackage{amssymb}
\usepackage{bm}
\usepackage{graphicx}
\def\bbbone{{\mathchoice {\rm 1\mskip-4mu l} {\rm 1\mskip-4mu l}
{\rm 1\mskip-4.5mu l} {\rm 1\mskip-5mu l}}}

\newcommand{\hf}{{\hat{\phi}}}
\newcommand{\hx}{\hat{x}}
\newcommand{\hk}{\hat{k}}
\newcommand{\hP}{\hat{\partial}}

\def\SO{\sf{SO}}
\def\he{{\hat e}}
\newcommand{\be}{\begin{equation}}
\newcommand{\ee}{\end{equation}}
\newcommand{\bea}{\begin{eqnarray}}
\newcommand{\eea}{\end{eqnarray}}

\begin{document}

\title{%
Field theory on $\kappa$--Minkowski space revisited: \\
Noether charges and breaking of Lorentz symmetry}
\author{Laurent Freidel}
\email{lfreidel@perimeterinstitute.ca} \affiliation{Perimeter
Institute, 31 Caroline Street North Waterloo, Ontario, Canada N2L
2Y5} \affiliation{Laboratoire de Physique, ENS Lyon, CNRS UMR 5672,
46 All\'ee d'Italie, 69364 Lyon Cedex 07}
\author{Jerzy Kowalski-Glikman}
\email{jurekk@ift.uni.wroc.pl} \affiliation{Institute for
Theoretical Physics, University of Wroc\l{}aw, Pl.\ Maxa Borna 9,
Pl--50-204 Wroc\l{}aw, Poland}
\author{Sebastian Nowak}
\email{snowak@ift.uni.wroc.pl} \affiliation{Institute for Theoretical Physics,
University of Wroc\l{}aw, Pl.\ Maxa Borna 9, Pl--50-204 Wroc\l{}aw, Poland}

\date{\today}
\begin{abstract}
This paper is devoted to detailed investigations of free scalar field theory on $\kappa$-Minkowski space. After reviewing necessary mathematical tools we discuss in depth the Lagrangian and solutions of field equations. We analyze the spacetime symmetries of the model and construct the conserved charges associated with translational and Lorentz symmetry. We show that the version of the theory usually studied  breaks Lorentz invariance in a subtle way: There  is an additional trans-Planckian mode
present, and an associated conserved charge (the number of such modes) is not a Lorentz scalar.
\end{abstract}

\maketitle

\section{Introduction}

The $\kappa$-Minkowski space \cite{kappaM1}, \cite{kappaM2} is
defined by the following Lie-type commutation
relation\footnote{Throughout this paper we choose the Planck units,
in which the Planck mass scale $\kappa$ as well as the Planck length
scale $1/\kappa$ are equal 1. }
\begin{equation}\label{1}
 [\hat x^0, \hat x^i] = i \hat x^i.
\end{equation}
As shown in \cite{kappaM1} this space can be constructed naturally
from $\kappa$-Poincar\'{e} algebra, and therefore is a natural
candidate for space-time in the context of Doubly Special Relativity
(see \cite{Amelino-Camelia:2000ge},\cite{Amelino-Camelia:2000mn},
\cite{jkgminl}, \cite{rbgacjkg} for original formulation of DSR and
\cite{Kowalski-Glikman:2004qa}, \cite{Kowalski-Glikman:2006vx} for
reviews.) It can be also related \cite{juse} to other DSR proposals,
like the Magueijo-Smolin model \cite{Magueijo:2001cr}.

There are strong indications that $\kappa$-Minkowski space arises also in the context of quantum gravity coupled to particles or fields. In the case of gravity in 3 spacetime dimensions some arguments has been presented in \cite{Amelino-Camelia:2003xp} and in \cite{Freidel:2003sp}. In the papers \cite{Freidel:2005bb}, \cite{Freidel:2005me}, \cite{Freidel:2005ec} it was shown that a field theory on non-commutative space arises directly from spin foam model of 3 dimensional gravity. Admittedly the non-commutative space differs in this case from (\ref{1}) but it can be argued \cite{Freidel:2003sp} and \cite{JurekBernd} that there must exist a 3d spin foam model leading to $\kappa$-Minkowski space directly. In the case of gravity in 4d there also exists a number of arguments indicating that in an appropriate ``no gravity limit'' $\kappa$-Minkowski space would emerge in effective description of fields, after integrating out  topological degrees of freedom of gravity that remain present even in this limit \cite{Kowalski-Glikman:2006mu}.

This motivates investigations of physics on
$\kappa$-Minkowski space, which can be analyzed by unraveling the properties of field theories
living on this space. This endeavor was undertaken by many authors
\cite{Kosinski:1999ix},\cite{Kosinski:2001ii},
\cite{Amelino-Camelia:2001fd}, \cite{Kosinski:2003xx},
\cite{Daszkiewicz:2004xy} (see also \cite{Dimitrijevic:2004nv}).
Unfortunately, our present understanding of this space is quite
incomplete. In turn this was a major stumbling block in improving
our understanding of DSR theories and their physical predictions.

In this paper we would like to report some progress that has been
achieved by using a mixture of methods employing both the $\kappa$-Minkowski space and  a complementary picture of field theory formulated on standard Minkowski spacetime, with an
appropriate star product replacing $\kappa$-Minkowski space fields
multiplication. It should be stressed that both these methods, mediated by a picture in which the field is defined on part of de Sitter momentum space \cite{Kowalski-Glikman:2002ft}, \cite{Kowalski-Glikman:2003we}, strongly rely on the structure of Borel group, whose Lie algebra provides a defining relation for $\kappa$-Minkowski space (\ref{1}). In \cite{LJSshort} we made use of the star product picture to construct a theory on Minkowski spacetime and to analyze some of its properties. Here we would like present an in depth analysis of field theory on $\kappa$-Minkowski space.

 Among the main results presented in this paper we give the explicit construction of all (translational and Lorentz)  conserved charges of a $\kappa$-Minkowski field theory associated with the Lorentz covariant calculus.
It turns out that the translational charges do satisfy the usual dispersion relation, which is in sharp contrast with the results of \cite{Agostini:2006nc},\cite{Arzano:2007gr} who computed the translational Noether charges for a non Lorentz covariant calculus.
We also show that there are two versions of the field theory on $\kappa$-Minkowski space, one which breaks explicitly Lorentz invariance and one which does not.
Both versions are associated with the same relativistically invariant lagrangian but differ in the range of integration over momentum space. The first one is the one usually studied  in the literature on
$\kappa$-Minkowski space and associated to a Hopf algebra construction.
The breaking of Lorentz symmetry is non trivial and implies that there is a conserved number of trans-Planckian (anti)particles. Each individual inertial observer sees a different number of such
modes.
The second is the one studied in \cite{LJSshort}  and shown to be equivalent to a non local field theory on usual Minkowski spacetime.

The plan of the paper is as follows. In the following section we recall some technical tools necessary for construction of the field theory on $\kappa$-Minkowski space.
We present the group structure and geometry underlying $\kappa$-Minkowski space and the structure of he covariant calculus. We also clarify the role of the Lorentz action on $\kappa$-Minkowski space and show how this action usually presented in a non linear form can be linearized by working with the right momentum variables. We summarize key results concerning the construction of the star product and finally study in great detail the structure of solutions of a free, massive kappa field theory.
 Section III is devoted to some remarks concerning bicovariant differential calculus and its relation to group theory. Section IV, which comprise the core of the paper contains discussion of lagrangian, field equations and their solutions, symmetries and conserved currents. In the next section we discuss symplectic structure of the theory and the algebra of charges. Section VI is devoted to discussion of our results. In the Appendix we collect some formulas that we frequently make use of in the main text.

\section{Preliminaries}

In this section we would like to collect the facts that will be necessary for construction of field theory on $\kappa$-Minkowski space, and for investigations of its properties. Most of the facts presented below are rather well known, some of them however are, to our knowledge, quite new. We start our discussion with momentum space.

\subsection{De Sitter space of momenta and group theory}

It is well know for quite some time that DSR and $\kappa$-Poincar\'e algebra are both closely related to the fact that the space of momenta is curved,
and has a form of elliptic  de Sitter space which is a $\mathbb{Z}_{2}$ identification of de Sitter space.
Elliptic de Sitter space of momenta can be thought of as a four dimensional hypersurface
\begin{equation}\label{2}
   -P_0^2 + P_1^2 + P_2^2 + P_3^2 + P_4^4 =1
\end{equation}
in five dimensional Minkowski space, modulo the identification $P_{A} = -P_{A}$.
On elliptic de Sitter space  one can introduce coordinates $(k_0, k_i)$ which cover only a half of de Sitter space
(\ref{2}) defined by $P_+ \equiv P_0 + P_4 = e^{k_0} >0$. In a cosmological setting these coordinates correspond to a flat slicing of deSitter space, they are defined as follows
\begin{eqnarray}
 {P_0}(k_0, \mathbf{k}) &=&  \sinh
{{k_0}} + \frac{\mathbf{k}^2}{2}\,
e^{  {k_0}}, \nonumber\\
 P_i(k_0, \mathbf{k}) &=&   k_i \, e^{  {k_0}}, \nonumber\\
 {P_4}(k_0, \mathbf{k}) &=&  \cosh
{{k_0}} - \frac{\mathbf{k}^2}{2}\, e^{  {k_0}}.
\label{3}
\end{eqnarray}
Another important choice of coordinates on elliptic de Sitter space (which do preserve Lorentz symmetry) is the one where we restrict to $P_{4}>0$, we'll make use of it in section \ref{starprod} while introducing the star product.
While de Sitter space is a homogeneous space $\SO(4,1)/\SO(3,1)$ the half of it covered by coordinates $(k_0, k_i)$ is in fact a Lie group, which we will call Borel group.
This  Borel group arises in the Iwasawa decomposition of the $\SO(4,1)$ \cite{Kowalski-Glikman:2004tz}.
The Lie algebra of the Borel group is generated by elements $\hat x^\mu\equiv J^{+ \mu}=  J^{4\mu}+J^{0\mu}$ where $J^{AB}$ denote the Lie algebra generators of  $\SO(4,1)$.
 The commutators are having the form
\begin{equation}\label{7}
 [\hat x^0, \hat x^i] = i \hat x^i
\end{equation}
and we use the convention $\hat x_{0}=-\hat x^{0},$ $\hat x_{i}=\hat x^{i}.$
Thus positions of $\kappa$-Minkowski space (\ref{1}) are nothing but generators of translations on de Sitters space of momenta (\ref{2}), as it should be.
 An explicit, five dimensional matrix representation of the generators of the algebra (\ref{7}) looks as follows
\begin{equation}\label{5drep}
\hat x^0 = -{i} \,\left(\begin{array}{ccc}
  0 & \mathbf{0} & 1 \\
  \mathbf{0} & \mathbf{0} & \mathbf{0} \\
  1 & \mathbf{0} & 0
\end{array}\right) \quad
\hat{\mathbf{x}} = {i} \,\left(\begin{array}{ccc}
  0 & {\bm{\epsilon}\,{}^T} &  0\\
  \bm{\epsilon} & \mathbf{0} & \bm{\epsilon} \\
  0 & -\bm{\epsilon}\,{}^T & 0
\end{array}\right),
\end{equation}
where $\bm{\epsilon}$ is a three dimensional vector with a single unit entry.

An ``ordered plane wave on $\kappa$-Minkowski space'' \cite{Amelino-Camelia:1999pm}
\begin{equation}\label{10}
   \he_k \equiv e^{ik_i \hat x^i} e^{ik_0 \hat x^0}
\end{equation}
has now a clear geometric interpretation of being a Borel group element.
The group structure being given by
\begin{equation}\label{group}
    \he_{kl} \equiv \he_k \he_l = e^{i\hat x^i(k_i + e^{-k_0}l_i)} e^{i \hat x^0(k_0 + l_0)}
\end{equation}
The composition of  plane waves can be equivalently described in terms of a non trivial  Hopf algebra structure for the momentum $k$.
 Since $k$ can be regarded as a function on Borel group, one can associate with it the non commutative coproduct dual to the group multiplication, which turns out to be
 \begin{equation}\label{14}
   \Delta(k_i) = k_i \otimes \bbbone + e^{-k_0} \otimes k_i, \quad \Delta(k_0) = k_0 \otimes \bbbone + \bbbone \otimes k_0
\end{equation}
Similarly the conjugate of a plane wave
\begin{equation}\label{15}
   (\he_k)^\dag = e^{-i  k_0 \hat{x}^0} e^{-i   k_i \hat{x}^i} = e^{-i \hat  (e^{k_0}k_i)\hat{x}^i}e^{-i  k_0 \hat{x}^0} = \he_{S(k)}
\end{equation}
gives the antipode
\begin{equation}\label{16}
   S(k_i) = - e^{k_0}k_i, \quad S(k_0) = - k_0.
\end{equation}
Knowing the coproduct and antipode for variables $k$ one can readily calculate the corresponding expressions for the momenta $P$, which are listed in the Appendix. They will turn out to be useful below.

 In will be convenient for us to know the  matrix representation of the Borel group element  which has the following form
\begin{equation}\label{11}
    \he_k = K_A{}^B= \left(\begin{array}{ccc}
  \bar P_4 & -\mathbf{P}e^{-k_0} & P_0 \\
  -\mathbf{P} & \mathbf{\bbbone} & -\mathbf{P} \\
  \bar P_0 & \mathbf{P}e^{-k_0} &  P_4
\end{array}\right)
\end{equation}
where $(P_0, \mathbf{P}, P_4)$ are given by (\ref{3}), while
\begin{eqnarray}
 \bar {P_4}(k_0, \mathbf{k}) &=&  \cosh
{{k_0}} + \frac{\mathbf{k}^2}{2}\,
e^{  {k_0}} \nonumber\\
\bar {P_0}(k_0, \mathbf{k}) &=&  \sinh
{{k_0}} - \frac{\mathbf{k}^2}{2}\, e^{  {k_0}}
\label{12}
\end{eqnarray}
Note that these additional momenta belong to an hyperbolic space $\bar{P}_{4}^{2} -\bar{P}^{2}_{0} -\bf{P}^{2} = 1$.

\subsection{Lorentz transformations}
It is well known that kappa-Minkowski non commutative space carries an action of a kappa-deformed version of the Poincar\'e algebra
\cite{kappaM1}. By duality this action becomes an action on the dual space of de Sitter momenta. Importantly but not surprisingly  this action is just the natural linear action of $SO(3,1)$ on de Sitter space which leaves $P_4$ invariant as we now show.

 Lorentz generators act on $\kappa$-Minkowski space coordinates in the standard way
$$
 M_{i}\triangleright \hat{x}_{0}=0,\;\;\; M_{i}\triangleright \hat{x}_{j}=i\epsilon_{ijk}\hat{x}_{k},
$$
 \begin{equation}\label{8a}
 N_{i}\triangleright \hat{x}_{0}=i\hat{x}_{i},\;\;\; N_{i}\triangleright \hx_{j}=i\delta_{ij}\hx_{0}.
\end{equation}
With the help of co-product this action can be extended to products of noncommutative
coordinates as follows
\begin{equation}\label{9a}
    N_i\triangleright (\hx\hat{y})=(N^1_i\triangleright \hx)(N^2_i\triangleright \hat{y})
\end{equation}
where we have used the Sweedler notation for co-product
$ \triangle t=\sum t^1_i\otimes t_i^2 $
and the coproducts of the Lorentz generators are given by
$$
\triangle (M_i)=M_i\otimes \bbbone+\bbbone\otimes M_i,
$$
\begin{equation}\label{s31}
    \triangle (N_i)=N_i\otimes \bbbone +e^{-{k_0}}\otimes N_i+\epsilon_{ijk}k_j\otimes M_k.
\end{equation}
One can check by direct calculation that the action of the Lorentz generators on a plane wave  is given by
\bea\label{Ni1}
 N_i\vartriangleright\hat{e}_k&=& i\, \left( {1\over 2} \left(1 -e^{-2{{k}_{0}}}\right) + {1\over 2} {\mathbf{k}}\,{}^{ 2}\, \right)  :
 \hat{x}_{i}\hat{e}_{k}:  - i\, {k}_{i} : \left({\mathbf{k}}{\mathbf{\hat{x}}} +\hat{x}_{0}\right)  \hat{e}_k: \\
 M_i\vartriangleright\hat{e}_k &=& i\epsilon^{ijk} k_{j}: \hat{x}_{k} \hat{e}_{k}:
 \eea
 where $:f(\hx):$ means ordered function with all $\hx_0$ shifted to
the right.

By moving $\hat{x}^{\mu}$ out of the normal ordering (\ref{Ni1}) we can simplify the action of
Lorentz transformations which then read
 \bea\label{Lor}
  N_i\vartriangleright\hat{e}_k&=&
 i\left(\hat{x}_i P_0(k)-\hat{x}_0P_i(k)\right) e^{-k_0} \hat{e}_k. \\
 M_i\vartriangleright\hat{e}_k&=& i\left(\epsilon^{ijk} P_{j}(k) \hat{x}_{k}\right) e^{-k_{0}}  \hat{e}_k
\eea
Let us introduce the derivative operators on momentum space as follows
\be
\nabla^{0}\equiv \frac{\partial}{\partial k_{0}}- k_{i}\frac{\partial}{\partial k_{i}},\quad \nabla^{0}\equiv \frac{\partial}{\partial k_{i}}.
\ee
It can be checked that these derivatives implement the right multiplication on the group, that is
$$ \nabla^{\mu} \he_{k} = i \hat{x}^{\mu} \he_{k}$$
and the generators of Lorentz transformation can be written
\be\label{Lor2}
  N_i\vartriangleright\hat{e}_k=
 e^{-k_0}\left( P_0(k)\nabla_{i}-P_i(k)\nabla_{0}\right)  \hat{e}_k, \quad
 M_i\vartriangleright\hat{e}_k = e^{-k_{0}}\left(\epsilon^{ijl} P_{j}(k) \nabla_{l}\right)   \hat{e}_k
\ee
One sees that the Lorentz  transformations acting on a function of $k$ are deformed and non linear, indeed
\begin{equation}\label{4}
  [M_i, k_j] = i\, \epsilon_{ijk} k_k, \quad [M_i, k_0] =0
\end{equation}
\begin{equation}\label{5}
   \left[N_{i}, {k}_{j}\right] = i\,  \delta_{ij}
 \left( {1\over 2} \left(
 1 -e^{-2{k_0}}
\right) + {{\mathbf{k}^2}\over 2}  \right) - i\, k_{i}k_{j} ,
\quad
  \left[N_{i},k_0\right] = i\, k_{i}.
\end{equation}
which are just the defining relations of $\kappa$-Poincar\'e algebra in the bicrossproduct basis \cite{kappaM1}.

It is possible  to linearize these transformation however if one writes them in terms of $P$.
In order to do so one first uses the chain rule $\nabla^{\mu}= \frac{\partial P_{A}}{\partial k_{\mu}} \partial_{P_{A}}$
and one finds that the right invariant derivatives  are indeed related to Lorentz generators
\be
\nabla_{0} = J_{+0}= -P_{0} \partial_{P_{4}} - P_{4}\partial_{P_{0}},\quad
\nabla_{i} = J_{+i} = P_{+} \partial_{P_{i}} - P_{i}(\partial_{P_{4}} -\partial_{P_{0}})
\ee
where $P_{+}(k)=P_{4}(k)+P_{0}(k)=e^{k_{0}}.$
Therefore the Lorentz transformation (\ref{Lor2}) are simply given by
\be\label{Lor3}
  N_i\vartriangleright\hat{e}_k=
 \left( P_0\partial_{P_i}-P_i\partial_{P_0}\right)  \hat{e}_k, \quad
 M_i\vartriangleright\hat{e}_k =  \left(\epsilon^{ijl} P_{j} \partial_{P_l}\right)   \hat{e}_k
\ee
and the non linear deformed commutation relations are mapped to the usual linear ones with  $P$  transforming as
a Lorentz vector
\be
[N_{i},P_{j}]= iP_{0}, \quad [N_{i},P_{0}]= iP_{i}.
\ee

\subsection{Differential calculus}

 To define the bicovariant differential calculus, the calculus covariant with respect to the Lorentz action, (see \cite{5dcalc1},
\cite{5dcalc2}, \cite{5dcalc3}), one considers the total differential
of a plane wave. Given any differential $d$ that satisfies Leibniz rules one can show that the action of $d$ on a plane wave is necessarily diagonal. Moreover
it is well known that a calculus on $\kappa$-Minkowski space, which is covariant with respect to Lorentz transformations must be at least five dimensional, so that
\begin{equation}\label{19}
 d \he_k = id\hat x^\mu\, \hat\partial_\mu \he_k +
  id\hat x^4\,\hat\partial_4 \he_k
  \equiv id\hat x^A\,\hat\partial_A \he_k
\end{equation}
We will discuss the covariance properties of the calculus in more details in Section III below.
The basic one-forms $d\hat x^A $ satisfy the following
commutation relations
\be\label{20aa}
 [\hat x^\mu,d\hat x^A] =  (x^{\mu})^{A}{}_{B} d\hat{x}^B,
\ee
where $(x^{\mu})^{A}{}_{B} $ denotes the matrix elements of $x^{\mu} $ in the $5d$ representation (\ref{5drep}).
With the help of Leibniz rule, that the differential $d$ (\ref{19}) satisfies by definition, one can check that\footnote{ Note that we have defined $\hat \partial_\mu$ to be $i$ times the usual derivative for future convenience}
\begin{equation}\label{24}
   \hat \partial_\mu \he_k = P_\mu\, \he_k, \quad \hat \partial_4 \he_k = (1-P_4)\, \he_k
\end{equation}
where $P_A$ are again given by (\ref{3}). Recall that $P_\mu$ transform in a standard linear way as components of a Lorentz vector, while $P_4$ is a Lorentz scalar. This  reflects the covariance of the calculus we use.
It
follows  that the differentials transform in the standard linear way under
action of rotations $M_{i} =\frac{i}2 \epsilon^{ijk}L_{jk}$ and boosts $N_{i}=iL_{0i}$
\be\label{21aa}
[L_{\mu\nu}, d\hat x^A] = \delta_{\mu}^{A} dx_{\nu}- \delta_{\nu}^{A} dx_{\mu}
\ee
which is the crucial property making the calculus Lorentz-covariant.

The last technical point to be mentioned concerns to properties of
differentials vis a vis conjugation introduced above (\ref{15}).
Using the identity
\begin{equation}\label{19a}
 d(\he_k \he_{S(k)} )=0
\end{equation}
we find the right derivative
\begin{equation}\label{20a}
 d\hf =-i\partial_A^\dagger\hf\,d\hat x_A
\end{equation}
where $\hf$ is any function which can be expressed as Fourier transform (\ref{fourier}) and
\begin{equation}\label{21a}
\hat{\partial}_\mu^\dagger \he_k(\hat{x})\equiv(\hat{\partial}_\mu
\he_k(\hat{x}))^\dagger=S(P)_\mu
\he_k(\hat{x}),\,\,\,\,\hP_4^\dagger=\hP_4
\end{equation}
With the help of this we find
\begin{equation}\label{22a}
  (d\hf)^\dagger=d\hf^\dagger.
\end{equation}

\subsection{Fields and Lorentz invariant action}
Given a ("time to the right ordered") field $\hat{\phi} = :\phi(\hat{x}):$ we define the translation invariant  integral to be
\be
\int_{\mathbb{R}^{4}} \hat{\phi}\equiv \int \mathrm{d}^{4}x\, \phi({x}).
\ee
This integral is the unique integral invariant under translation
\be
\int_{\mathbb{R}^{4}} \hat{k}_{\mu}\vartriangleright  \hat{\phi}=0.
\ee

Two remarks concerning this integral are in order.  First it is not a cyclic since
\be
\int_{\mathbb{R}^{4}} \he_{k}\he_{p} = \delta(k_{0}+p_{0})\delta^{3}(\mathbf{k}+e^{-k_{0}}\mathbf{p})
= e^{3k_{0}} \delta(p_{0}+k_{0})\delta^{3}(\mathbf{p}+e^{-p_{0}}\mathbf{k})
= e^{3k_{0}}\int_{\mathbb{R}^{4}} \he_{p}\he_{k}
\ee
However it satisfy the exchange property
\be
\int_{\mathbb{R}^{4}}\he_{k}^{\dagger} \he_{p} =\int_{\mathbb{R}^{4}}\he_{p}^{\dagger} \he_{k}
\ee
and this property extends to functions, which can be expressed as Fourier integrals.

Second, if one wants to have an integral invariant only under the covariant calculus  defined above, the integral is no longer unique
since any function independent of $x_{i}$ and  such that $f(x_{0}+i)=f(x_{0})$ is constant for this calculus.
To see this let us consider  the function $\hat{f} \equiv f({\zeta} ^{2})$ where  $f$ is arbitrary and ${\zeta}=e^{\pi \hat{x}^{0}}$.
 Then from the definition of the differential calculus one can see that
  $\hat{\partial}_{A} \hat{f} = 0$ and we can define the translation invariant integral
  $\int_{f}  \hat{\phi}\equiv \int_{\mathbb{R}^{4}} \hat{f}(\zeta^{2}) \hat{\phi}.$
  In order to get rid of this ambiguity we will now assume that a function constant for the covariant calculus is really constant that is
  we will quotient the non commutative algebra of function on $\kappa$-Minkowski space by the relation relation $\zeta^{2}=1$.
  It is possible to define such a quotient since $\zeta^{2}$ commutes with any $\hat{f}$.  From now on we will assume that this relation is implemented.

Using this integral we can define the Fourier coefficients  and the inverse Fourier transform\footnote{The proof of the inversion formula goes schematically as follows
\bea
\int_{\mathbb{R}^{4}} \he_{S(k)} \left(\int_{B}\mathrm{d}\mu(p) \,\, \he_{p} \tilde{\phi}(p)\right)= \int_{\mathbb{R}^{4}} \int_{B}\mathrm{d}\mu(p) \,\, \he_{S(k)p} \tilde{\phi}(p)=
 \int_{B}\mathrm{d}\mu(p)\left( \int_{\mathbb{R}^{4}} \,\, \he_{p}\right) \tilde{\phi}(k p) =\tilde{\phi}(k)
\eea}
 to be
\be \label{fourier}
\tilde{\phi}(k)= \int_{\mathbb{R}^{4}} \he_{S(k)} \hat{\phi}, \quad\quad  \hat{\phi} =\int_{B}\mathrm{d}\mu(k) \,\, \he_{k} \tilde{\phi}(k)
\ee
where $B$ denotes the Borel group $\mathrm{d}\mu(k)=\frac{e^{3k_{0}}}{(2\pi)^{4}}\mathrm{d}{k_{0}} \mathrm{d}^{3}{\mathbf{k}}$ is the left invariant measure on it,
$\mathrm{d}\mu(pk)=\mathrm{d}\mu(k)$.

The conjugation of plane waves extends directly to conjugation of fields, to wit
\begin{equation}\label{18}
     \hf^\dagger(\hat{x})=\int \mathrm{d}\mu(k)  {\tilde\phi}^{*}(k)\,\he_{S(k)}
\end{equation}
where $*$ denotes the standard complex conjugation.

We will be interested in a free massive scalar theory, given by the Lorentz invariant  Lagrangian\footnote{Since $P_{4}$ is  Lorentz scalar some authors take as a Lagrangian
$ \hat {\cal L} = \hf^\dag (\hP_{4}- M) \hf $ where $M=1+\sqrt{1+m^{2}}$. This choice is less natural from our point of view and can be obtained from the case we study by restricting the solutions to the one satisfying $P_{4}>0$. }

\begin{equation}\label{41}
   \hat {\cal L} =\frac12\left[ (\hP_\mu \hf)^\dag \hP^\mu \hf + m^2\hf^\dag \hf\right]
\end{equation}
which leads to  the equation of motion
$$
\hP_{\mu}\hP^\mu \hf + m^2 \hf=0.
$$
The action can be expressed in terms of Fourier modes as follows
\be\label{action}
 S= \int_{\mathbb{R}^{4}} \hat {\cal L} = \int \mathrm{d}\mu(k) \tilde{\phi}^{*}(k)\left(P^{\mu}P_{\mu}(k) + m^{2}\right)\tilde{\phi}(k).
\ee

\subsection{More on plane waves and fields}
Using the isomorphism between Borel group and the half of de Sitter defined by $P_{+}>0$ we can label plane waves
by points in half of de Sitter space. We denote the corresponding plane wave by $\he_{P}$, with  $\he_{P(k)}\equiv \he_{k}$
by definition.
We can extend the definition of a non commutative plane wave to all of de Sitter space by defining
\be
\he_{-P(k)}\equiv  \he_{(k_{0}+i\pi,\mathbf{k})}=\he_{k}\zeta
\ee
These plane waves are diagonal under the action of the covariant calculus
\be \hat{\partial}_{A} \he_{P} = P_{A} \he_{P} \ee
The general product of two non commutative plane wave is given by
\be
\he_{P} \he_{Q} = \he_{P\oplus Q}.
\ee
where we have $P_{+} \equiv P_{4} + P_{0} $, $P_{-} \equiv P_{4} - P_{0} = \frac{1-\mathbf{P}^{2}}{P_{+}}$  and
\bea
(P\oplus Q)_0&=&  P_{0}Q_{+}+ \frac{Q_{0}}{P_{+}} + \frac{\mathbf{P}\cdot \mathbf{Q}}{P_{+}} , \,\,\,\,\,\,\quad  \quad
(P\oplus Q)_i = P_{i} Q_{+} + Q_{i}, \,\,\,\,\,\, \quad \quad
(P\oplus Q)_+= P_{+}Q_{+}.
\eea
we also have  $\he_{P}^{\dagger}=\he_{S(P)}(x)$
where
\be
S(P)_{0} =  -P_{0} + \frac{\mathbf{P}^{2}}{P_{+}}, \, \quad
S(P)_{i} =  - \frac{P_{i}}{P_{+}}, \quad S(P_{+})= P_{+}^{-1} .
\ee
Note that the product and antipodes  are defined whenever $P_{+}\neq 0$ but there is no need to restrict to $P_{+}>0$.

Making the change of variables from $k$ to $P$
a general field on kappa-Minkowski space defined in (\ref{fourier}) can be written in terms of
an integral on de Sitter space \cite{LJSshort}
\be\label{mode}
\hat{\phi}  =\int \frac{d^{5}P}{(2\pi^{4})} \theta(P_{+})\delta(P_{A}P^{A}-1) \tilde{\phi}(P) \he_{P}
\ee
 where abusing the notation slightly we have denoted $\tilde{\phi}(P(k))= \tilde{\phi}(k)$, and the Heaviside function imposes the constraint $P_{+}>0$.
  Similarly the action (\ref{action}) can be written
  \be\label{action2}
 S = \int \frac{d^{5}P}{(2\pi^{4})} \theta(P_{+})  \tilde{\phi}^{*}(k)\left(P^{\mu}P_{\mu}(k) + m^{2}\right)\tilde{\phi}(k).
\ee
Note that we have explicitly included the restriction $P_{+} \equiv P_0 + P_4= e^{k_0}>0$ since this is on this sector covering half of de Sitter space that the theory and the non commutative  product and were initially defined. This restriction explicitly breaks Lorentz invariance  (since the condition $P_+=P_0+P_4=0$ is is not preserved by boosts).

The only way to cover half de Sitter space while preserving Lorentz invariance is to choose $P_{4}>0$ instead as a restriction. This suggests that if one wants to produce a relativistic invariant field one just has to insert
$\theta(P_{4})$ instead of $\theta(P_{+})$ in the mode expansion (\ref{mode}).
Unlike the restriction on $P_{+}$ however, the condition $P_{4}>0$ is not preserved by the non commutative multiplication and fields with such condition would not form an algebra. One way around is to suppose that the field is a field on full de sitter space which is  even under the $\mathbb{Z}_{2}$ identification $\tilde{\phi}_{R}(P)=\tilde{\phi}_{R}(-P)$ so that it describe a function on elliptic de Sitter space, and we use the product of elliptic de Sitter space. This is detailed in the next sections.

The relativistic invariant  field is then given by
\be
\hf_{R} = \frac12\int \frac{d^{5}P}{(2\pi)^{4}}\theta(P_{4})\delta(P_{A}P^{A}-1)\tilde{\phi}_{R}(P) \he_{P}
\ee

\subsection{Star product}
\label{starprod}

In order to  represent the previous action in the language of effective field theory
as a non local action which can be expanded in terms of higher order derivative operators we need to express
the non commutative structure of kappa-Minkowski space in terms of a star product.
A star product  is chosen once we give a Weyl  mapping
$\mathcal{W}$ from plane waves $\he_P$ on the
non-commutative space described earlier to a function on
ordinary Minkowski space-time, with coordinates $x_\mu$.
Here we restrict to the case where $\he_P$ is mapped to a plane wave.
The Weyl map should be invertible, which means that from the knowledge of $P$ one should be able to reconstruct uniquely a point in elliptic de Sitter hence  a
real 4-vector $(k_{0},k_{i})$ and we want this map to be Lorentz covariant. Thus we postulate
\begin{equation}\label{25}
 \mathcal{W}\left(  \he_{k}\right) =e^{i\tilde{P}(k)_\mu \, x^\mu}\equiv E_{\tilde{P}}(x),
\end{equation}
where $\tilde{P}_{\mu} =k_{\mu} + O(P)$.

This defines the star product $\star$, to wit
\begin{equation}\label{26}
{\mathcal{W}}\left(  \he_{k} \he_{p} \right) \equiv e^{i\tilde{P}(k)_\mu \,
x^\mu}\star e^{i\tilde{P}(l)_\mu \, x^\mu} = e^{i\tilde{P}(kp)_{\mu}x^{\mu}}=W\left(  \he_{kp} \right)
\end{equation}
Obviously there are as many such star products as the are functions
$\tilde{P}(k)$. This is a huge ambiguity and it is related, in the context of DSR,  to the ambiguity of
the choice of basis of the momentum space.
Note however that this ambiguity does not change the form of the action (\ref{action}) and should not change the physics,
Changing the form of the Weyl map amounts to a non local field redefinition.
Indeed given two Weyl maps $\mathcal{W}_{1}, \mathcal{W}_{2}$ the corresponding fields are related by
$$ \mathcal{W}_{1}(\hat\phi)(x) = e^{i((\tilde{P}_{1}\circ \tilde{P}_{2}^{-1}(-i\partial ))_{\mu} +i\partial_{\mu})x^{\mu}} \mathcal{W}_{2}(\hat \phi)(x)$$

Among all possible choice one is preferred from the point of view of Lorentz covariance this is the choice for which
the Lorentz covariant derivative $\hat{\partial}_{\mu}$  on
$\kappa$-Minkowski space introduced above is mapped by $\mathcal{W}$ to the the standard
derivative on Minkowski space as follows
\begin{equation}\label{27aa}
  \mathcal{W} \left(\hat\partial_\nu  \he_{P} \right) =\frac1i\,\partial_\nu
E_{P}(x)= P_\nu \, E_{P}(x)
\end{equation}
This is the choice made in \cite{LJSshort}.

Let us pause for a moment to discuss the Weyl map (\ref{27aa}) in more details.
It should be noted that in general the knowledge of $\hat{\partial}_{\mu}$ does not determine the knowledge of the value of $\hat{\partial}_{4}$. This means in particular that the coordinates $P_0, \ldots, P_3$ of the points in region $0$ are the same as the coordinates of some points in region $+$ (see Figure 1). One possible solution of this problem is as follows. In order to have an invertible Weyl map one need to identify  elliptic de Sitter space with the portion of de Sitter covered by $P_{4}=\sqrt{1-P^{2}}>0$, $P^{2}=P_{0}^{2}-\mathbf{P}^{2}$.
The restriction to positive $P_{4}$ makes the Weyl map  invertible and covariant.
This means that $$ \mathcal{W}(\he_{k}) = E_{\epsilon({k}) P(k)}$$ where $P(k)$ is given by (\ref{3}) and
$\epsilon(k) =-1 $ if $P_{4}(k)<0$ that is if $ k^{2}> 1 +e^{-2k_{0}}$ and $\epsilon(k) =1 $
otherwise.  This map is invertible with inverse given by
$$ \mathcal{W}^{-1}(E_{P}) = \he_{\epsilon_{P}(P_{\mu}, \sqrt{1+P_{0}^{2}- \mathbf{P}^{2}})} =  \he_{\epsilon_{P}P}$$ where
$\epsilon_{P} =-1$ if $P_{+}\equiv P_{0} + \sqrt{1+P^{2}} <0 $ and
$\epsilon_{P} =+1$ otherwise.
The composition of momenta is the same as the one described in the previous section as long as all the momenta involved are small, in general the composition involve signs and cocycles which insures that the composition is well defined.

From the group product one can straightforwardly deduce that
\be\label{star}
\left(E_{P}\star E_{Q}\right)(x) = E_{\left(\epsilon_{(P,Q)}(\epsilon_{P}P)\oplus (\epsilon_{Q}Q) \right)}(x)
\ee
where  the deformed addition is defined in the previous section and $\epsilon_{(P(k),Q(p))}= \epsilon(kp)=\pm1$ is a cocycle.
we also have  $E_{P}^{\dagger}(x) =E_{S(P)}(x)$

Using the expansion (\ref{fourier}) in Fourier modes the star product can be defined on
all functions on $\kappa$-Minkowski spacetime.
\bea
\phi(x)\equiv \mathcal{W}(\hat{\phi})(x) &=& \int_{B}\mathrm{d}\mu(k) \,\, \mathcal{W}(\he_{k}) \tilde{\phi}(k) \\
&=& \int \frac{\mathrm{d}^{5}P}{(2\pi)^{4}} \delta(P_{A}P^{A}-1) \theta(P_{+})  \,  E_{P}(x) \tilde{\phi}(P)\label{FoP1}
\eea
Where we have expressed the integral over the Borel Group in terms of an integral over
half of de Sitter space \cite{LJSshort}.

Since $P(k)=0$ iff $k=0$ when $k\in \mathbb{R}^{4}$ and $\mathrm{det}(\frac{\partial{P}}{{\partial k}})|_{k=0}=1 $ the integral over
kappa-Minkowski space is mapped by the Weyl map to the usual integral on $\mathbb{R}^{4}$
\be
\int_{\mathbb{R}^{4}} \hat{\phi}
=\int \mathrm{d}^{4} x \,\,W(\hat{\phi})(x).
\ee
and the Fourier transformation can be written \cite{LJSshort}
\be
\tilde{\phi}(P) =  \int  \mathrm{d}^{4} x \,\, (E^{\dagger}_{P} \star \phi)(x) =|P_{4}| \int \mathrm{d}^{4} x \,\,E^{*}_{P}(x)  \phi(x)
\ee
where $*$ denotes the complex conjugation, $E^{*}_{P}(x)= e^{-iP \cdot x}$.

 We can now  derive the action of Lorentz symmetry generators on functions defined with the help of ${ W}$-map on Minkowski spacetime. From (\ref{Lor}) we know how boost generator $N_i$ acts on a $\kappa$-Minkowski plane wave. By applying the Weyl map we get
\begin{equation}\label{35a}
N_i \triangleright \phi(x) \equiv  {\cal W}( N_i \triangleright \hf(\hx))= (x_i \star \partial_0 - x_0\star \partial_i) e^{-k_0} \phi \end{equation}
With the help of identities obtained by taking derivatives of (\ref{star})
$$
x_i \star \phi(x) = (x_i e^{k_0} - x_0 \partial_i)\phi(x), \quad x_0 \star \phi(x) = (x_0 e^{k_0} - x_0 \partial_0)\phi(x)
$$
one finds that
\begin{equation}\label{35b}
N_i \triangleright \phi(x) = (x_i  \partial_0 - x_0 \partial_i) \phi(x) \end{equation}
so as promised the boost generator action on Minkowski spacetime fields is just the standard one. Analogously one can check that the action of rotations is standard as well.

It is crucial to note that even if the action on the field is the usual one, once we use the Weyl map, the action
(\ref{action})
\be S= \frac12 \int \mathrm{d}^{4} x \,\,\left[\left((\partial_{\mu}\phi)^{\dagger} \star \partial^{\mu}\phi \right)(x) - m^{2} (\phi^{\dagger} \star \phi )(x)\right]
\ee
 is {\it not} Lorentz invariant.
This is because the measure
$\mathrm{d}\mu(k)=\frac{\mathrm{d}^{5}P}{(2\pi)^{4}} \delta(P_{A}P^{A}-1) \theta(P_{+})$ contains a  restriction
$P_{+}>0$ which is not preserved by  boost transformations.

One can remedy to this problem if one interpret the Fourier transform $\tilde{\phi}(P_{A})$ (defined so far only on the sector $P_{+}>0$) as  a function on elliptic de Sitter space, that is as an even function on de Sitter space  $\tilde{\phi}(P_{A})=\tilde{\phi}(-P_{A})$
 for all value of $P_{+}$.
Given this function we can construct a relativistic invariant field
\be
\phi_{R}(x)\equiv  \int \frac{\mathrm{d}^{5}P}{(2\pi)^{4}} \delta(P_{A}P^{A}-1) \theta(P_{4})  \,  E_{P}(x) \tilde{\phi}(P)\label{FoP}
\ee
and a Lorentz invariant action
\be S_{R}= \frac12 \int \mathrm{d}^{4} x \,\,\left[\left((\partial_{\mu}\phi_{R})^{\dagger} \star \partial^{\mu}\phi_{R} \right)(x) - m^{2} (\phi_{R}^{\dagger} \star \phi_{R} )(x)\right]
\ee
As shown in \cite{LJSshort} this action is equivalent to a non local action defined on Minkowski spacetime
\be S_{R}= \frac12 \int \mathrm{d}^{4} x \,\, \left[(\partial_{\mu}\phi_{R})^{*}(x) \sqrt{1+\Box} \, \partial^{\mu}\phi_{R}(x)  - m^{2} \phi_{R}^{*}(x) \sqrt{1+\Box}\, \phi_{R}(x) \right]
\ee
where $\Box = -\partial_{0}^{2} +\partial_{i}^{2}$.
We are now left with two version of kappa Poincar\'e field theory; one non relativistic which is the version usually referred to as a kappa field theory and one relativistic which was studied in \cite{LJSshort}. We now study both but with more emphasis on the former version.

\subsection{Solutions of equations of motion}
\begin{figure}
\includegraphics[angle=0,width=13cm]{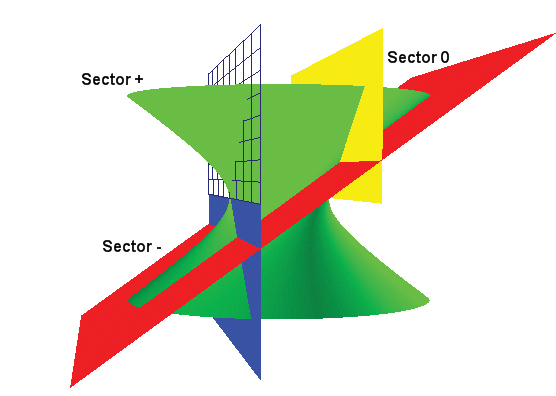}
\caption{The momentum space is the portion of De Sitter space above the plane $P_+=0$ where $P_{0}$ is the vertical axis. The mass shell is given by the intersection of this portion of de Sitter space with the vertical planes $P_{4}=\pm cste$. This mass shell naturally decomposes into three sectors indicated: $+$ with positive energy and $P_4>0$, $-$ with negative energy and $P_4>0$, and $0$ with  positive energy and $P_4<0$. Notice that in the limit $\kappa\rightarrow\infty$ the second sector becomes unbounded, while the third sector disappears.}
\end{figure}


We are now interested in the solutions of the equation of motion
\be\label{eom}
\hP_{\mu}\hP^\mu \hf + m^2 \hf=0.
\ee
A general solution of the equation (\ref{eom}) can be written in terms of Fourier modes
$$
\hat{\phi}  =\int \frac{d^{5}P}{(2\pi^{4})} \theta(P_{+})\delta(P_{A}P^{A}-1)\delta(P_{\mu}P^{\mu}+ m^{2})\tilde{\phi}(P) \he_{P}
$$

Note that we have explicitly included the restriction $P_{+} \equiv P_0 + P_4= e^{k_0}>0$ since this is on this sector covering half of de Sitter space that the theory and the non commutative  product and were initially defined.
As stressed above this restriction explicitly breaks Lorentz invariance, and the relativistically invariant field one may choose to work with is given by
$$
\hf_{R}=\int d^{5}P \theta(P_{4})\delta(P_{A}P^{A}-1)\delta(P_{\mu}P^{\mu}+ m^{2})\tilde{\phi}_{R}(P) \he_{P}$$
it is related to the previous one by the following transformation
\be
\tilde{\phi}_{R}(P_{A})=\tilde\phi(-P_{A}), \, \mathrm{if} \, P_{0}<0, P^{2}>1,\quad \tilde{\phi}_{R}(P_{A})=\tilde\phi(P_{A})
\,\,\mathrm{otherwise}.
\ee
From now on we will work with the field $\hf$ since in that case the coproduct rule and antipode are the usual ones.
The previous transformation will allow us to map all the results obtained for $\hf$ in terms of  $\hf_{R}$.

We can solve the delta constraints which a priori give four sectors depending on whether
$P_{0}$ and $P_{4}$ are positive or negative. The constraint $P_{+}=P_{0}+P_{4}>0$ eliminates one sector $P_{4}<0,P_{0}<0$, we are thus left with
three types of solutions (see Figure 1). In sharp difference from the usual case,where there are only two sectors. This three sectors are denoted $+,0,-$ and are such that
\bea
+ &:& P_{0} = +\omega_{\mathbf{P}}, P_{4} = +\sqrt{1+m^2} \\
- &:& P_{0} = -\omega_{\mathbf{P}}, P_{4} = +\sqrt{1+m^2}, \bf{P}^{2}<1\\
0 &:& P_{0} = +\omega_{\mathbf{P}}, P_{4} = -\sqrt{1+m^2}, \bf{P}^{2}>1
\eea
with
$$\omega_{\mathbf{P}} =\sqrt{\mathbf{P}^{2}+m^{2}}.$$
Thus, decomposing the field $\hf$ into modes belonging to these three sectors we find
\begin{equation}\label{phidecomp}
    \hf= \int\frac{d^3P}{2\omega_{\mathbf{P}}
|P_4|}a_{+}(\mathbf{P})\he^{+}_{\mathbf{P}}+\int_{|\bf{P}|<1}\frac{d^3P}{2\omega_{\mathbf{P}}
|P_4|}a_{-}(\mathbf{\mathbf{P}})\he^{-}_{\mathbf{P}}
+\int_{|\bf{P}|>1}\frac{d^3P}{2\omega_{\mathbf{P}}
|P_4|}a_{0}(\mathbf{\mathbf{P}})\he^{0}_{\mathbf{P}}
\end{equation}
where $$\he^{\epsilon}_{\mathbf{P}}\equiv \he_{\left(P_{0}(\epsilon),\mathbf{P}(\epsilon),P_{4}(\epsilon)\right)},\quad \epsilon = +,-,0$$
and we have  introduced the notation
$$P_{0}(+)=-P_{0}(-)= P_{0}(0)=\omega_{\mathbf{P}},\quad P_{i}(+)=-P_{i}(-)= P_{i}(0)={P}_{i}\quad P_{4}(+)= P_{4}(-)=-P_{4}(0)=\sqrt{1+m^{2}}.$$

Note in passing that the relativistic field $\hf_{R}$ is obtained by changing the 5-momenta of the $0$ sector $P_{A}(0)\to -P_{A}(0)$ while living invariant the
sectors $+,-$. $\hf_{R} $ is a scalar if
\bea
a_{+}(\Lambda\mathbf{P})=a_{+}(\mathbf{P}),\quad &a_{-}(\Lambda\mathbf{P})=a_{-}(\mathbf{P}),\, |\mathbf{P}|\leq1,\,|\Lambda \mathbf{P}|\leq1 \\
\quad a_{0}(\Lambda\mathbf{P})=a_{0}(\mathbf{P}) \,|\mathbf{P}|\geq1,\, |\Lambda \mathbf{P}|\geq1
&\quad a_{0}(\Lambda\mathbf{P})=a_{-}(\mathbf{P})\,|\mathbf{P}|\leq1,\, |\Lambda \mathbf{P}|\geq1.
\eea

Our conjugation involves the antipode which is given  by
$$
S(P)_{i}= -\frac{P_i}{P_{4}+ P_{0}},\quad
S(P)_{0}=-P_{0}+\frac{\mathbf{P}^2}{P_{0}+P_{4}}=-\frac{m^2+P_{0}P_{4}}{P_{0}+P_{4}} ,\quad S(P_{4})=P_{4}.
$$
It is important to note that the antipode exchanges the sectors $+$ with $-$ and maps $0$ onto itself.

In each sector it is a function of $\mathbf{P}$, we define
  $S^{+}_{\mathbf{P}}\equiv -S(P)$, $P\in [+]$, $S^{-}_{\mathbf{P}}\equiv S(-P)$, $-P\in [-]$,
$S^{0}_{\mathbf{P}}\equiv S(P)$, $P\in [0]$, explicitly
\bea
 \left(S^{+}_{\mathbf{P}}\right)_{0}= \frac{m^{2}+\sqrt{1+m^{2}}\omega_{\mathbf{P}}}{\sqrt{1+m^{2}}+\omega_{\mathbf{P}}}
\quad \left(S^{+}_{\mathbf{P}}\right)_{i}= \frac{P_{i}}{\sqrt{1+m^{2}}+\omega_{\mathbf{P}}} \\
\left(S^{-}_{\mathbf{P}}\right)_{0}= \frac{\sqrt{1+m^{2}}\omega_{\mathbf{P}}-m^{2}}{\sqrt{1+m^{2}}-\omega_{\mathbf{P}}}
\quad \left(S^{-}_{\mathbf{P}}\right)_{i}= \frac{P_{i}}{\sqrt{1+m^{2}}-\omega_{\mathbf{P}}}\\
 \left(S^{0}_{\mathbf{P}}\right)_{0}= \frac{ \sqrt{1+m^{2}}\omega_{\mathbf{P}}-m^{2}}{\omega_{\mathbf{P}}-\sqrt{1+m^{2}}}
\quad \left(S^{0}_{\mathbf{P}}\right)_{i}= \frac{-P_{i}}{\omega_{\mathbf{P}}-\sqrt{1+m^{2}}}
 \eea
from this we can see  that
 \be
(S^{\epsilon}_{\mathbf{P}})_{0} = \omega_{\mathbf{S^{\epsilon}_{P}}}, \,\, \epsilon=\pm,0.
\ee
and that we also have
\be
\mathbf{S^{-}_{S^{+}_P}= P},\quad \mathbf{S^{+}_{S^{-}_P}= P}, \, \mathbf{P}^{2}<1,\quad \mathbf{S^{0}_{S^{0}_P}= P}, \, \mathbf{P}^{2}>1.
\ee

In order to perform the conjugation we will need to change variables $\mathbf{P}\to \mathbf{S^{\epsilon}_{P}}$.
Under this change of variable the measure transform as
\be
\mathrm{d}^{3}S^{\epsilon}_{\mathbf{P}} = \mathrm{d}^{3}\mathbf{P} \mathrm{det}(\partial_{P_{i}}(\mathbf{S^{\epsilon}_{P}})_{j}))=
\frac{\mathrm{d}^{3} \mathbf{P}} {|P_{+}(\epsilon)|^{3}} \frac{\omega_{\mathbf{S^{\epsilon}_{P}}} }{\omega_{\mathbf{P}} }
\ee
with $P_{+}(\epsilon)=  P_{0}(\epsilon)+P_{4}(\epsilon)$.
Thus the conjugate field is given by
\bea\label{e15}
\hf^\dagger&=&
\int\frac{d^3P}{2\omega_{\mathbf{P}} |P_4|}a_{+}^{*}(\mathbf{P})\he^{-}_{S^{+}_{\mathbf{P}}}+\int_{|P|<1}\frac{d^3P}{2\omega_{\mathbf{P}} |P_4|}{a_{-}^{*}(\mathbf{P})}\he^{+}_{S^{-}_{\mathbf{P}}}+\int_{|P|>1}\frac{d^3P}{2\omega_{\mathbf{P}} |P_4|}a_{0}^{*}(\mathbf{P})\he^{0}_{S^{0}_{\mathbf{P}}} \nonumber\\
&=&
\int\frac{d^3P}{2\omega_{\mathbf{P}} |P_4|}a_{-}^{\dagger}(\mathbf{P})\he^{+}_{\mathbf{P}}+
\int_{|P|<1}\frac{d^3P}{2\omega_{\mathbf{P}} |P_4|}{a_{+}^{\dagger}(\mathbf{P})}\he^{-}_{\mathbf{P}}+\int_{|P|>1}\frac{d^3P}{2\omega_{\mathbf{P}} |P_4|}a_{0}^{\dagger}(\mathbf{P})\he^{0}_{\mathbf{P}}\label{pdagform}
\eea
where
\bea
a_{-}^{\dagger}(\mathbf{P}) \equiv \frac{a_{-}^{*}(\mathbf{S^{+}_{P}})}{|P_{+}(+)|^{3}}, \quad
a_{+}^{\dagger}(\mathbf{P}) \equiv \frac{a_{+}^{*}(\mathbf{S^{-}_{P}})}{|P_{+}(-)|^{3}}, \quad
a_{0}^{\dagger}(\mathbf{P}) \equiv  \frac{a_{0}^{*}(\mathbf{S^{0}_{P}})}{|P_{+}(0)|^{3}}.
\eea
One sees that positively charged particles are conjugate to negatively charge particles of bounded momenta $\mathbf{P}^{2}<1$ whereas the
trans-Planckian particles of type $0$  are self conjugate.

In our calculation of conserved charges associated with symmetries of the theory it will be convenient to use the field $\hf$ (\ref{phidecomp}) and its conjugate (\ref{pdagform}) written in the following compact notation
\begin{equation}\label{phicompact}
   \hf = \sum_\epsilon\int_\epsilon\, \frac{d^3P}{2\omega_{\mathbf{P}} |P_4|}a_{-\epsilon}^{\dagger\,*}(\mathbf{P})\he^{-\epsilon}_{\mathbf{S}^\epsilon_\mathbf{P}}, \quad \hf^\dag = \sum_\epsilon\int_\epsilon\, \frac{d^3P}{2\omega_{\mathbf{P}} |P_4|}a_{-\epsilon}^{\dagger}(\mathbf{P})\he^{\epsilon}_{\mathbf{P}}.
\end{equation}

\section{More on bicovariant differential calculus}

Let us start this section with somehow different view on the bicovariant differential calculus introduced above. Usually this calculus is derived by making use of an abstract algebraic theory that obscures somehow it physical meaning and properties. Here we will utilize  the already stressed relation between $\kappa$-Minkowski space, $\kappa$-Poincar\'e algebra and group theory of Borel group and algebra.

Let us start with a slight change of notation. The coordinate forms $d\hat x^\mu$ together with the form $d\hat x^4$ we will denote collectively by $\epsilon^A$. $\epsilon^A$ form a basis of the space of forms, right invariant under translation. As it was shown above (\ref{20aa}), (\ref{21aa}) they form a representation of $\kappa$-Minkowski space algebra and thus of Borel group. In terms of the plane waves it reads
\begin{equation}\label{36}
   \he_k\, \epsilon^A \he_k^{-1} = \epsilon^B\, K_B {}^A
\end{equation}
where $K_B {}^A$ is a group element in five dimensional representation, explicitly given by (\ref{11}).

We define the differential to be
\begin{equation}\label{37}
    d = \epsilon^A\, \hat \partial_A
\end{equation}
and we demand that it satisfies Leibniz rule, i.e.,
$$
d \he_{kl} = d\he_k\, \he_l + \he_k\,d \he_l
$$
In terms of derivatives this can be written as
\begin{equation}\label{38}
   \epsilon^A\, \hat \partial_A \he_{kl} = \epsilon^A\, \hat \partial_A\he_k\, \he_l + \he_k\,\epsilon^A\he_k^{-1}\, \he_k\hat \partial_A \he_l = \epsilon^A \left(\hat \partial_A\he_k\, \he_l + K_A{}^B\, \he_k\,\hat \partial_B \he_l\right)
\end{equation}
Using the fact that while acting on product of functions (using Sweedler notation) $\hat \partial_A(\hf\hat\psi) = \sum (\hat \partial^{(1)}_A\hf)(\hat \partial^{(2)}_A\hat \psi)$ this last expression can be understood as expressing coproduct rule for derivative
\begin{equation}\label{39}
   \Delta(\hat \partial_A) =\hat \partial_A\otimes \bbbone +  K_A{}^B \otimes \hat \partial_B
\end{equation}
or, by replacing $\hat \partial_\mu$ with $P_\mu$ and $\hat \partial_4$ with $1-P_4$, as a corresponding coproduct for momenta. Let us check this explicitly for $P_i$ (or what is the same for $\hat \partial_i$.) Indeed, multiplying the matrix $K_A{}^B$ (\ref{11}) by the derivatives vector, and replacing derivatives with corresponding momenta we find
$$
\Delta(P_i) = P_i\otimes \bbbone + P_i \otimes \left( P_0 + (P_4-\bbbone)\right) + \bbbone\otimes P_i = P_i \otimes \left( P_0 + P_4\right) + \bbbone\otimes P_i
$$
which is nothing but (\ref{a.4}). Two other coproducts (\ref{a.5}), (\ref{a.6}) can be recovered analogously.

In the construction above we have been considering the differential associated with translations. However translations are not the only transformations associated with the group action on $\kappa$-Minkowski space, Lorentz transformations play an equally important role in the game. Thus it seems natural to extend the differential so that it includes the latter as well. Let us therefore instead of (\ref{37}) consider
\begin{equation}\label{40}
    d_F = \epsilon^A\, \hat \partial_A + \omega^{\mu \nu}\, \hat x_\mu \, \hat \partial_\nu e^{-k_0} = \epsilon^A\, \hat \partial_A + \omega^{\mu \nu}\, L_{\mu \nu}
\end{equation}
This differential must again satisfy Leibniz rule $d_F(\hf \hat \psi) = d_F(\hf) \hat \psi +\hf d_F(\hat \psi)$. Since this condition is linear, by the considerations above we can consider just the Lorentz part of (\ref{40}). As in (\ref{38}) we get
\begin{equation}\label{40a}
   \omega^{\mu \nu}\, L_{\mu \nu} \he_{kl} = \omega^{\mu \nu}\, L_{\mu \nu}\he_k\, \he_l + \he_k\,\omega^{\mu \nu}\,  \he_k^{-1}\, \he_k L_{\mu \nu} \he_l = \omega^{\mu \nu} \left(L_{\mu \nu}\he_k\, \he_l + K_{\mu\nu}{}^{\rho \sigma}\, \he_k\,L_{\rho \sigma} \he_l\right)
\end{equation}
As in the derivation for translations above this expression must be consistent with the coproduct structure. Let us start therefore with the different end and derive the form of the matrix $K_{\mu\nu}{}^{\rho \sigma}$. The co products for rotations and boosts read (cf.\ Appendix)
$$
\triangle (M_i)=M_i\otimes \bbbone+\delta_i{}^j\otimes M_j
$$
$$
    \triangle (N_i)=N_i\otimes \bbbone +e^{-{k_0}}\delta_i{}^j\otimes N_j+\varkappa_i{}^j\otimes M_j
$$
where $\varkappa_i{}^j=\epsilon_{ik}{}^jk_k$. These equations can be written together as
$$
\Delta \left(
         \begin{array}{c}
           M \\ N
         \end{array}
       \right) =\left( \begin{array}{c}
           M \\ N
         \end{array}
       \right) \otimes \bbbone + \left(
                                   \begin{array}{cc}
                                     1 & 0 \\
                                     \varkappa & e^{-k_0} \\
                                   \end{array}
                                 \right)\otimes\left( \begin{array}{c}
           M \\ N
         \end{array}
       \right)
$$
Turning now to the generators $L_{\mu\nu}$ one can easily check that
\begin{equation}\label{41b}
   \Delta L_{\mu\nu} = L_{\mu\nu} \otimes \bbbone + \delta_{[\mu}{}^\rho\, k_{\nu]}{}^\sigma \otimes L_{\rho\sigma}
\end{equation}
with
$$
k_\nu{}^\sigma = \left(
                   \begin{array}{cc}
                     2e^{-k_0}-1 & 2\mathbf{k} \\
                     0 & \bbbone  \\
                   \end{array}
                 \right)
$$
In this way we find an explicit form of the matrix $K$ in (\ref{40a})
\begin{equation}\label{41c}
    K_{\mu\nu}{}^{\rho \sigma} = \delta_{[\mu}{}^\rho\, k_{\nu]}{}^\sigma
\end{equation}

It is natural to assume that the differential $d_F$ (\ref{40}) is nilpotent $d^2_F=0$. Denoting collectively $\epsilon^\alpha = (\epsilon^A, \omega^{\mu\nu})$ and $\hP_\alpha = (\hP_A, L_{\mu\nu})$ one can easily check that the condition of nilpotency is equivalent to the following requirement
\begin{equation}\label{41d}
   \hP_\alpha\, \epsilon^\beta = f^\beta_{\alpha\gamma}\, \epsilon^\gamma
\end{equation}
where $f^\beta_{\alpha\gamma}$ are structure constant of the algebra of $\hP_\alpha$ (which is in our case just the standard Poincar\'e algebra with the additional element $\hP_4$, which commutes both with translations and Lorentz transformations, see below.) Notice that it follows from this condition that
\begin{equation}\label{41e}
   \hP_A \, \epsilon^B =  \hP_A \, \omega^{\mu\nu} = 0
\end{equation}
which will be important in our calculation of conserved current and charges below, and that $ \epsilon^\mu = dx^\mu$ transform under Lorentz as components of vectors while $\epsilon^4$ is a Lorentz scalar.

The origin of the presence of the additional element $\partial_4$ and the corresponding parameter $\epsilon^4$, although proved  in \cite{5dcalc1}, \cite{5dcalc2} was somehow obscured, but it can be easily understood in our present framework. To see this it is sufficient to realize that the consistency of our procedure require that
$$
d_F (\he_k \epsilon^\alpha \he_k^{-1}) = d_F(\epsilon^\beta K_\beta{}^\alpha)
$$
But in the limit of infinitesimal $k$, (i.e., when the group element is can be approximated by $1$ + algebra element) $\he_k$ becomes just a unit plus a Lie algebra element $k \hat x$, so that in the parenthesis on the left hand side we have just one plus commutator, while $K_\beta{}^\alpha$ on the right hand side becomes a unit matrix plus a constant one $k_\mu\, \gamma_{\beta}{}^{\mu\alpha}$. Thus the above equation can be equivalently written as
\begin{equation}\label{41f}
    d_F ([x^\mu, \epsilon^\alpha]) = d_F \, \epsilon^\beta \, \gamma_{\beta}{}^{\mu\alpha}
\end{equation}
We will not solve this equation explicitly, because this is exactly the equation that has appeared as a key requirement in the analysis presented in \cite{5dcalc1}, \cite{5dcalc2}, with the result that the minimal translational sector must be five dimensional. This justifies the choice made in this paper.

It would be interesting to see explicitly what goes wrong with the four dimensional calculus, for example the one with the only nontrivial commutator being
\begin{equation}\label{41g}
    [x^0, dx^i] = i dx^i
\end{equation}
with
$$
N_i \triangleright dx_j = \delta_{ij}\, dx_0, \quad  N_i \triangleright dx_0 = dx_i
$$
It is worth stressing in passing that there is no ambiguity in the last equation  since the calculus must satisfy $N \triangleright df \equiv d( N \triangleright f)$. If one now applies $N_i$ to both sides of (\ref{41g}), with $N$ action on product defined by coproduct, of course, everything is consistent, and the formula  (\ref{41g}) is covariant. This result is not hard to understand because (\ref{41g}) from the point of view of Lorentz action is as covariant as the defining commutator (\ref{1}). However if one applies $N$ to the commutator $[x^i, dx^j] =0$ one finds that the result on the left hand side is non-zero \cite{5dcalc1}, which means that this commutator is not Lorentz-covariant and makes the four dimensional calculus incompatible with Lorentz symmetry.

\section{Conserved charges for translations and Lorentz transformation}

In this section we will construct the conserved charges associated with translational and Lorentz symmetries of the free scalar field Lagrangian
 \begin{equation}\label{41a}
   \hat {\cal L} =\frac12\left[ (\hP_\mu \hf)^\dag \hP^\mu \hf - m^2\hf^\dag \hf\right]
\end{equation}
Since the steps of  this construction are quite delicate, our presentation will be very detailed.
The construction of conserved charge for translational symmetry and for the non lorentz covariant 4-dimensional differential calculus have been first construct in \cite{Agostini:2006nc}, \cite{Arzano:2007gr}. Our derivation here is related but more general. First we consider the 5 dimensional calculus and as a conclusion we find that the conserved charge do satisfy indeed the usual dispersion relation in contrary to the result of  \cite{Agostini:2006nc}. Also we find that there are not 4 but 5 conserved charges. The fifth conserved charge is in the relativistic version of kappa-Poincar\'e field theory just the $U(1)$ charge however in the standard non relativistic version it has as we will see the  interpretation of the number of trans-Planckian particles.

\subsection{Generalities}

To calculate the charges associated with symmetries we must first decompose variation of the Lagrangian (\ref{41a})
into total derivative and the term proportional to field equations. To do that we make use of the formulas
$$
\hP_0(\hf \hat{\psi})= (\hP_0\hf)(e^{\hk_0}\hat{\psi}) +  (e^{-\hk_0}\hf)(\hP_0\hat{\psi})+(e^{-\hk_0}\hP_i\hf)(\hP_i\hat{\psi})$$ $$
\hP_i(\hf \hat{\psi})= (\hP_i\hf)(e^{\hk_0}\hat{\psi}) +  \hf(\hP_i\hat{\psi}), \quad
e^{\hk_0}(\hf \hat{\psi})=
(e^{k_0}\hf)(e^{\hk_0}\hat{\psi})
$$
 Using these
deformed Leibnitz rules satisfied by the derivatives one gets
\bea (\hP_{i}
\hf)^\dagger (\hP_{i} \delta\hf) &=& \hP_{i} \left ( (\hP_{i}
\hf)^\dagger \delta\hf \right) -
\hP_{i}(\hP_{i} \hf)^\dagger  e^{\hk_{0}} \delta\hf \nonumber\\
&=& \hP_{i}\left( (\hP_{i} \hf)^\dagger  \delta\hf \right) - e^{\hk_{0}} \left(  e^{-\hk_{0}}\hP_{i}(\hP_{i} \hf)^\dagger \delta\hf \right) \nonumber\\
&=& \hP_{i}\left( (\hP_{i} \hf)^\dagger  \delta\hf \right) +
e^{\hk_{0}} \left(  ( \mathbf{\hP}^2 \hf)^\dagger \delta\hf \right) \eea
Similarly  the term involving time
derivative gives

\bea (\hP_{0} \hf)^\dagger (\hP_{0} \delta\hf) &=&
\hP_{0} \left (  e^{\hk_{0}}(\hP_{0} \hf)^\dagger  \delta\hf \right)
-
 \left (  e^{\hk_{0}}\hP_{0}(\hP_{0} \hf)^\dagger  e^{\hk_{0}}\delta\hf \right)
-\hP_{i}(\hP_{0} \hf)^\dagger  \hP_{i} \delta\hf  \nonumber \\
&=& \hP_{0} \left (  e^{\hk_{0}}(\hP_{0} \hf)^\dagger  \delta\hf
\right) - e^{\hk_{0}}\left (  \hP_{0}(\hP_{0} \hf)^\dagger
\delta\hf \right) -\hP_{i}\left ( \hP_{i}(\hP_{0} \hf)^\dagger
\delta\hf \right) + \mathbf{\hP}^{2} (\hP_{0} \hf)^\dagger
e^{\hk_{0}}\delta\hf
\nonumber \\
&=& \hP_{0} \left (  e^{\hk_{0}}(\hP_{0} \hf)^\dagger  \delta\hf
\right) -\hP_{i}\left (  \hP_{i}(\hP_{0} \hf)^\dagger  \delta\hf
\right) - e^{\hk_{0}}\left ( (\hP_{0} -
e^{-\hk_{0}}\mathbf{\hP}^{2})(\hP_{0}
\hf)^\dagger  \delta\hf \right)\nonumber\\
&=& \hP_{0} \left (  (e^{-\hk_{0}}\hP_{0} \hf)^\dagger  \delta\hf
\right) +\hP_{i}\left (  (\hP_{i}e^{-\hk_{0}}\hP_{0} \hf)^\dagger
\delta\hf \right) + e^{\hk_{0}}\left ( (\hP_{0}^{2} \hf)^\dagger
\delta\hf \right)
\eea
Finally the variation of the mass
term gives
\bea m^{2}\phi^\dagger \delta \phi&=& -(e^{\hk_{0}}-1)
\left(m^{2}\phi^\dagger \delta \phi\right)
+e^{\hk_{0}}\left(m^{2}\phi^\dagger \delta
\phi\right)\nonumber\\
&=&  -(\hP_{0} -\hP_{4}) \left(m^{2}\phi^\dagger \delta \phi\right)
+e^{\hk_{0}}\left(m^{2}\phi^\dagger \delta
\phi\right)\nonumber\\
&=&   -\hP_{0}\left(m^{2}\phi^\dagger \delta \phi\right)
+\hP_{4}\left(m^{2}\phi^\dagger \delta \phi\right)
+e^{\hk_{0}}\left(m^{2}\phi^\dagger \delta \phi\right)
\eea
 With the help of these formulas, for the lagrangian (\ref{41}) we find
 \be\label{var} \delta {\hat{\cal L}} = \hP_{A}
\left(\hat{\Pi}^A \delta \hf\right) + e^{\hk_{0}}\left((\hP_{\mu}\hP^\mu \hf + m^2
\hf)^\dagger \delta\hf \right) + \mathrm{h.c} \ee with canonical momenta being defined as follows \bea
-\hat{\Pi}^{0}=\hat{\Pi}_0 &\equiv& \left(e^{-\hk_{0}}\hP_{0} \hf +
m^{2}\hf\right)^\dagger \label{PI0},\\
\hat{\Pi}^i=\hat{\Pi}_i &\equiv&
\left(\hP_{i}(1-e^{-\hk_{0}}\hP_{0})\hf\right)^\dagger,\\
\hat{\Pi}^4=\hat{\Pi}_4 &\equiv& \left(m^{2}\hf\right)^\dagger. \eea
It is worth noticing that although the zero component of field momentum (\ref{PI0}) looks quite strange, by using the definition of conjugated derivatives
\be\label{partialdag}
\hP_i^\dagger =-e^{-\hk_0}\hP_i, \quad \hP_0^\dagger=-\hP_0 +
\mathbf{\hP}^2e^{-\hk_0}, \hP_4^\dagger = \hP_4, \quad \left(e^{\hk_0}\right)^{\dagger}= e^{-\hk_0}
\ee
one can easily check that
\begin{equation}\label{Pi0}
    \hat{\Pi}_0=\hP_4\, \hP_0 \hf^\dag
\end{equation}
which means that on shell it differs from the standard time derivative of the field just by a constant multiplicative factor $\sqrt{1+m^2}$.

Because of the
Leibnitz rule, for $\delta\phi=d\phi$ we have:
$$
\hP_{A} \left(\hat{\Pi}^A d \hf\right)+\hP_{A}^\dagger \left( (d
\hf)^\dagger\hat{\Pi}^{\dagger A}\right)-d\hat{\cal L}=0
$$
In the first term the differential is placed to the right of the canonical momenta $\Pi$, but this can be easily corrected by noticing that the differential $d$ satisfies Leibniz rule by definition, so that
\begin{equation}\label{varLd}
   \hP_{A} \left(d(\hat{\Pi}^A  \hf)-d\hat{\Pi}^A  \hf\right)+\hP_{A}^\dagger \left( (d
\hf)^\dagger\hat{\Pi}^{\dagger A}\right)-d\hat{\cal L}=0
\end{equation}
Notice that in the formula above we could substitute the generalized differential $d_F$ for $d$, defined in Section III since they both satisfy Leibniz rule. We will make use of this below, when calculating charges associated with Lorentz symmetry. Before turning to this let us compute explicitly the translational charges.

\subsection{Energy-momentum tensor and conserved momenta}

Taking $d = \epsilon^a \hP_A$ and using the covariance identity $\hat\partial_{A} \epsilon^{B} =0$ proven earlier and discarding $\epsilon$ we find
$$
\hP_{A} \left(\hP_B(\hat{\Pi}^A  \hf)-\hP_B\hat{\Pi}^A
\hf\right)+\hP_{A}^\dagger \left( \hP_B
\hf^\dagger\hat{\Pi}^{\dagger A}\right)-\hP_B\hat{\cal L}=0
$$
or
\begin{equation}\label{conserveq}
    -\hP_{A} \left(\hP_B\hat{\Pi}^A \hf\right) +
\hP_{A}^\dagger \left( \hP_B\hf^\dagger\hat{\Pi}^{\dagger A}\right)
+\hP_B \left( \hP_{A}(\hat{\Pi}^A  \hf)- \hat{\cal L}\right)=0
\end{equation}
Notice that using (\ref{var}) with $\delta\hf =\hf$, $\delta\hf^{\dagger}=0$
one finds that for a free field  $\hP_{A}(\hat{\Pi}^A  \hf)- \hat{\cal L}=0$, so that last term drops out.

Using this fact, with the help of formulas for the conjugated derivatives (\ref{partialdag})
we can convert them to  the usual ones and  write the conservation (\ref{conserveq}) equation as
$$
\hP_{A} T^{A}{}_{B}=0
$$
where the components of the energy momentum tensor have the following form
\bea
T^{0}{}_{B}&=&-\hP_B\hat{\Pi}^0 \hf -\hP_{B}\hf^{\dagger} \Pi^{0\dagger} \\
T^{i}{}_{B}&=&-\hP_B\hat{\Pi}^i \hf - e^{{-k_{0}}}(\hP_{B}\hf^{\dagger} \Pi^{i \dagger})
+e^{{-k_{0}}}\hP^{i}(\hP_{B}\hf^{\dagger} \Pi^{0 \dagger})  \\
T^{4}{}_{B}&=&-\hP_B\hat{\Pi}^4 \hf +\hP_{B}\hf^{\dagger}\Pi^{4\dagger}
= 0
\eea
where in the last equation we use the explicit expression of $\Pi^{4}$.
Thus we have five $4$-dimensional conservation equations.
\begin{equation}\label{pT=0}
    \hat{\partial}_\mu{T}^\mu_B=0.
\end{equation}

Let us note in passing that in the case of an interacting field
we would still have
$$
\hP_{A}\left(T^{A}_{B} + \delta_{B}^{A}(\hP_{A}(\hat{\Pi}^A  \hf)- \hat{\cal L})\right)=0.
$$
 We could then define a new current in the following way
$$
\tilde{T}^\mu_B=T^\mu_B+\delta^{\mu}_{A}(\hP_{A}(\hat{\Pi}^A  \hf)- \hat{\cal L})
+\hat{\partial}^\mu(2-\hat{\partial}_4)^{-1}(T^4_B +\delta_{\mu}^{A}(\hP_{A}(\hat{\Pi}^A  \hf)- \hat{\cal L}))=0
$$
and since
$$
\hat{\partial}_4=\hat{\partial}_\mu\hat{\partial}^\mu(2-\hat{\partial}_4)^{-1}
$$
the following $4$ dimensional conservation equation holds
$$
\hat{\partial}_\mu\tilde{T}^\mu_B=0.
$$

Returning to the free case, we just have shown that  the conserved charges are
$$
\mathcal{P}_B=\int_{\mathbb{R}^{3}}  T_B^0 = -\int_{\mathbb{R}^{3} } (\hP_B\hat{\Pi}^0\hf+\hP_B\hf^\dagger\hat{\Pi}^{\dagger 0}).
$$
Before we start the calculation we must choose one of the possible field expansions. Since the differential operators in the formula above hits the first term, it is most convenient to use for $\hf^\dag$ the formula in the second line of (\ref{pdagform}) with an appropriate formula for $\hf$, so that there is no antipode in the plane waves in expansion of $\hf^\dag$. Thus we will use the field expansion in the compact form (\ref{phicompact}).

The integral over $\mathbb{R}^{3}$ is defined as the using integral on  Minkowski space, with the help of ${\cal W}$ map, with the help of the following identity
 $$
{\cal{W}}\left(\int_{\mathbb{R}^{3} }  \hat{e}_{p} \hat{e}_{k}^{\dagger}\right)= \int d^3 x e^{iP_\mu x^\mu}\star e^{iS(K)_\mu x^\mu}
= e^{i( \frac{P_{0}}{ K_+}-\frac{K_{0}}{P_+})x^0}  \delta^3\left(\frac{\mathbf{P}-\mathbf{K}}{K_{+}}\right)
$$
\begin{equation}\label{calc1}
= e^{i\left(\frac{P_{0}}{ K_+}-\frac{K_{0}}{P_+}\right)x^0} |{K_{+}}|^{3} \delta^3\left({\mathbf{P}-\mathbf{K}}\right)
\end{equation}
Using this formula along with (\ref{Pi0}) we can express the charge ${\cal W}({\cal{P}}_{A})$ as a momentum space integral
\begin{equation}\label{calc2}
-\sum_{\epsilon, \epsilon'}\int_{\epsilon, \epsilon'}
\frac{d^3P\, d^3K}{4\omega_{\mathbf{P}}\omega_{\mathbf{K}} |P_4||K_4|} P_{A}(\epsilon) \left( P_0(\epsilon) P_4(\epsilon) + K_0(\epsilon') K_4(\epsilon') \right) a_{-\epsilon}^{\dagger}(\mathbf{P}) a_{-\epsilon'}^{\dagger *}(\mathbf{K})
e^{i\left(\frac{P_{0}}{ K_+}-\frac{K_{0}}{P_+}\right)x^0}|{K_{+}(\epsilon')}|^{3}\delta^3(\mathbf{P}-\mathbf{K})
\end{equation}
where the subscripts $\epsilon, \epsilon'$ of the integral label the range of integration of the different sectors.
Now it is easy to see that the integral vanishes for
 $\epsilon\neq \epsilon'$ and that the nonzero contributions with $\epsilon= \epsilon'$ are time independent, as it should be since
by construction the charge is conserved, but it is reassuring to check it directly.
Therefore we get
\bea
{\cal{P}}_{A} &=&
 \sum_{\epsilon} \int_{\epsilon} \frac{d^3\mathbf{P} }{ 2\omega_{\mathbf{P}} |P_4|}
\alpha(\epsilon)  P_{A}(\epsilon)  |{P_{+}(\epsilon)}|^{3} a_{-\epsilon}^{\dagger}(\mathbf{P})a_{-\epsilon}^{\dagger *}(\mathbf{P}) \\
&=&
 \sum_{\epsilon} \int_{\epsilon} \frac{d^3\mathbf{P} }{ 2\omega_{\mathbf{P}} |P_4|}
\alpha(\epsilon)  P_{A}(\epsilon)   a_{-\epsilon}^{\dagger}(\mathbf{P})a_{-\epsilon}^{}(\mathbf{S^{\epsilon}_P}) \label{B.1} 
\eea
where $\alpha(\epsilon)=\mathrm{sgn}(P_{4}(\epsilon)P_{0}{(\epsilon)})$, is the U(1) charge
$$\alpha(+)=+1,\quad \alpha({-})=-1,\quad\alpha({0})=-1.$$ Explicitly the $U(1)$ conserved charge is
$${\cal Q}=-\int (\Pi^{0}\hf -\hf^{\dagger}\Pi^{0\dagger})= \sum_{\epsilon} \int \frac{d^3\mathbf{P} }{ 2\omega_{\mathbf{P}} |P_4|}
\alpha(\epsilon)N_{{\epsilon}}(\mathbf{P})$$
Here we introduced the particle number  $N_{{\epsilon}}(\mathbf{P})=  a_{-\epsilon}^{\dagger}(\mathbf{P})a_{-\epsilon}^{}(\mathbf{S^{\epsilon}_P})$.
Now we can write more explicitly the conserved charges associated with translational symmetry.
\bea
{\cal{P}}_{0} &=&
\int_{\epsilon} \frac{d^3\mathbf{P} }{ 2\omega_{\mathbf{P}} |P_4|}
 \left( N_{+}(\mathbf{P}) + N_{-}(\mathbf{P}) - N_{0}(\mathbf{P})\right) \omega_{\mathbf{P}} \\
{\cal{P}}_{4} &=&
- \int_{\epsilon} \frac{d^3\mathbf{P} }{ 2\omega_{\mathbf{P}} }
 \left( N_{+}(\mathbf{P}) - N_{-}(\mathbf{P}) + N_{0}(\mathbf{P})\right) \\
 {\cal{P}}_{i} &=&
 \int_{\epsilon} \frac{d^3\mathbf{P} }{ 2\omega_{\mathbf{P}}|P_4| }
 \left( N_{+}(\mathbf{P}) - N_{-}(\mathbf{P}) + N_{0}(\mathbf{P})\right)P_{i}
 \eea
  Equivalently, we can
write them in terms of the antipode
\bea{\cal{P}}_{A} &=&   \sum_{\epsilon} \int_{\epsilon} \frac{d^3\mathbf{P}}{2\omega_{\mathbf{P}}|P_4|}
 \alpha(-\epsilon) \left(S^{\epsilon}_{\mathbf{P}}\right)_{A}  a_{\epsilon}^{*}(\mathbf{P})a_{\epsilon}(\mathbf{P}).
\eea

Since each mode contributes the energy $\omega_{\mathbf{P}} \equiv \sqrt{m^2 + \mathbf{P}^2}$ and the momentum $\mathbf{P}$ to the total conserved energy and momentum, respectively, we see that indeed (in the quantum field theory language) for a single particle state the standard dispersion relation $P_0^2 - \mathbf{P}^2 =m^2$ holds. The issue of multiparticle states and their properties will be discussed in a separate paper.

Looking at the above formulas it seems that we are having a problem since the particle of type $0$ have negative energy.
However  the number of particle of type $0$ is also conserved since their number can be expressed as a combination of conserved charges introduced above $-2{\cal{N}}_{0}= \sqrt{1+m^{2}}{\cal Q}+ {\cal P}_{4}$ and
therefore no instability occurs.

So remarkably the theory behave in that respect, and for the sector of type $0$, much like a non relativistic theory: each inertial observer will see a fixed number of particle of type $0$ while type $+,-$ particles behave in a relativistic manner.
Moreover different inertial observers will see a different number of type $0$ particle.
Indeed, under a boost positive energy particle of type $-$ with $\mathbf{P}^{2}<1$ will be converted to particle of type $0$  with $\mathbf{P}^{2}<1$.
For the boosted observer the number of particle of type 0 will be different from the not boosted one but still be conserved.
If one starts from a state where all particle have momenta below the Planck scale the theory will be Lorentz invariant until
the boost parameter is high enough as to allow for trans-Planckian momenta, in this case type $-$ particle will be converted to type 0.

The charges for the relativistic theory are then easily deduced from this, $P_{A}(0)\to -P_{A}(0)$
essentially amount for the momentum charge to the change  $N_{0}(\mathbf{P})\to -N_{0}(\mathbf{P})$
that is
\bea
{\cal{P}}_{0} &=&
\int_{\epsilon} \frac{d^3\mathbf{P} }{ 2\omega_{\mathbf{P}} |P_4|}
 \left( N_{+}(\mathbf{P}) + N_{-}(\mathbf{P}) + N_{0}(\mathbf{P})\right) \omega_{\mathbf{P}} \\
{\cal{P}}_{4} &=&
- \int_{\epsilon} \frac{d^3\mathbf{P} }{ 2\omega_{\mathbf{P}} }
 \left( N_{+}(\mathbf{P}) - N_{-}(\mathbf{P}) - N_{0}(\mathbf{P})\right) \\
 {\cal{P}}_{i} &=&
\int_{\epsilon} \frac{d^3\mathbf{P} }{ 2\omega_{\mathbf{P}}|P_4| }
 \left( N_{+}(\mathbf{P}) - N_{-}(\mathbf{P}) - N_{0}(\mathbf{P})\right)P_{i}
 \eea

Therefore one sees that we have only positive energy particles in this case, moreover the charge $\mathcal{P}_{4}$ is equal to the $U(1)$ charge $\mathcal{Q}$.
It follows that the number of particles of type $0$ is no longer conserved, and they can freely turn to the particles of type $-$ and vice versa. Now we see that combine particles of type $-$ and $0$ into one species, and the boundary between sub- and trans-Planckian modes disappears.

\subsection{The charges associated with Lorentz transformations}

 To complete this section let us finally calculate the conserved charges resulting from the invariance with respect to Lorentz transformations. The starting point will be equation (\ref{varLd}) applied to the Lorentz part of differential $d_F$ (\ref{40}), so that in this equation we substitute $\omega^{\mu \nu}\, L_{\mu\nu}= \omega^{\mu \nu}\, \hat x_\mu \, \hat \partial_\nu e^{-k_0}$ for $d$.

Using the fact that, as shown in Section III $\hP_A \omega^{\mu \nu} =0$, (\ref{41e}), we can discard it to find
$$
\hP_{A} \left(L_{\mu\nu}(\hat{\Pi}^A  \hf)-L_{\mu\nu}\hat{\Pi}^A
\hf\right)+\hP_{A}^\dagger \left( L_{\mu\nu}
\hf^\dagger\hat{\Pi}^{\dagger A}\right)-L_{\mu\nu}\hat{\cal L}=0
$$
As in the case of translations we would like to get rid of the first term by making use of the field equations. Here however changing order of derivative and Lorentz generator will produce an additional term, to wit
$$
-\hP_{A} \left(L_{\mu\nu}\hat{\Pi}^A \hf\right) +
\hP_{A}^\dagger \left( L_{\mu\nu}\hf^\dagger\hat{\Pi}^{\dagger A}\right) + [\hP_{A}, L_{\mu\nu}] \, \hat{\Pi}^A \hf
+L_{\mu\nu} \left( \hP_{A}(\hat{\Pi}^A  \hf)- \hat{\cal L}\right)=0
$$
The term in the parenthesis vanishes on shell, as before and since $[\hP_{A}, L_{\mu\nu}]$ is proportional to derivative we again have conservation equation for the resulting current ${\cal L}^A_{\mu\nu}$. The commutator contributes to the conserved charge only in the case of boost, where it contains term $\hP_0 g_{Ai}$ with $g$ being the Minkowski metric.

Repeating the reasoning of subsection B, and going to the Minkowski space time with the help of Weyl map, one easily finds that the conserved charge associated with Lorentz transformations looks very similar to the standard case and reads
\begin{equation}\label{C.1}
   {\cal L}_{\alpha\beta} = - \int x_{[\alpha} \star e^{-k_0}\,  T^0_{\beta]} + \delta_{\alpha0} \delta_\beta^i\int \partial_i \phi
\end{equation}
As usual we can omit the second term by taking appropriate boundary conditions for the field $\phi$.
The factor $e^{-k_0}$ seems somehow odd at the first sight, but it is exactly the one required to turn from the star product $x \star$ to the standard multiplication, cf.\ the discussion in Subsection II F and (\ref{35a}), (\ref{35b}).

In the case of the charge associated with space rotations ${\cal M}_{ij}$, going to the momentum space, replacing $x$ with derivative over $P$, and repeating the steps that has led to formula (\ref{B.1}) in the case of translational charge,  we find that the formula analogous to (\ref{calc2}) takes the form
\begin{equation}\label{C.2}
 \sum_{\epsilon, \epsilon'}\int_{\epsilon, \epsilon'}
\frac{d^3P\, d^3K}{4\omega_{\mathbf{P}}\omega_{\mathbf{K}} |P_4||K_4|}
 \left( P_0(\epsilon) P_4(\epsilon) + K_0(\epsilon') K_4(\epsilon') \right) a_{-\epsilon}^{\dagger}(\mathbf{P}) a_{-\epsilon'}^{\dagger *}(\mathbf{K})
e^{i\left(P\oplus S(K)\right)_{0} x^0}|{K_{+}(\epsilon')}|^{3}\, {K_{+}(\epsilon')}\, P_{[i}(\epsilon) \frac{\partial}{\partial P^{j]}}\delta^3(\mathbf{P}-\mathbf{K})
\end{equation}
where
\be\label{spk}
\left(P\oplus S(K)\right)_{0}= \frac{P_{0}}{ K_+}-\frac{K_{0}}{P_+}-\frac{K_i}{P_+}\, \frac{P_i-K_i}{K_+}
\ee
Now we can integrate by parts and then, noticing that there is no contribution from the boundary terms (since the both boundaries of sectors $-$ and $0$ are defined by $|P|=1$ and are rotationally invariant), we can integrate delta out to obtain
\begin{equation}\label{C.3}
   {\cal{M}}_{ij} =\frac1i\,
 \sum_{\epsilon} \int_{\epsilon} \frac{d^3\mathbf{P} }{ 2\omega_{\mathbf{P}} |P_4|}
\alpha(\epsilon)   |{P_{+}(\epsilon)}|^{3} P_{[j}(\epsilon)  \left( \frac\partial{\partial P^{i]} }\, a_{-\epsilon}^{\dagger}(\mathbf{P})\right) a_{-\epsilon}^{\dagger *}(\mathbf{P})
\end{equation}
By virtue of the same argument as in the previous subsection we see that there are no ``cross-sector'' contributions and that the time dependence in exponent drops out.

Let us now turn to the charge associated with boost, which can be read off from (\ref{C.1}) and using (\ref{35b}) to be
\begin{eqnarray}\label{C.4}
  {\cal N}_{i} &=&  \int\left( x_{0} \star e^{-k_0}\,  T^0_{i} - x_i \star e^{-k_0}\, T^0_{0}\right)
 =\int\left( x_{0} \left(   T^0_{i} -  e^{-k_0}\,  [\partial_{0}T^0_{i} - \partial_{i} T^{0}_{0}]\right) - x_i T^0_{0}\right) \\
 &=&- t {\cal P}_i - \int x_i \,  T^0_{0 }
\end{eqnarray}

In expressing the second term in (\ref{C.4}) in the form analogous to that in eq.\ (\ref{C.2}) one has to be a bit more careful not to forget to include all the terms in the time part of the plane wave (the terms that disappeared as a result of integrating out the momentum delta will contribute now, as we will see). Remembering about that we find
\begin{eqnarray}
     \frac1i\, \sum_{\epsilon, \epsilon'}\int_{\epsilon, \epsilon'}&&\!\!
\frac{d^3P\, d^3K}{4\omega_{\mathbf{P}}\omega_{\mathbf{K}} |P_4||K_4|}  P_0(\epsilon)  \left( P_0(\epsilon) P_4(\epsilon) + K_0(\epsilon') K_4(\epsilon') \right) a_{-\epsilon}^{\dagger}(\mathbf{P}) a_{-\epsilon'}^{\dagger *}(\mathbf{K}) \times\nonumber\\
     && \exp\left[i\left(\frac{P_{0}}{ K_+}-\frac{K_{0}}{P_+}-\frac{K_i}{P_+}\, \frac{P_i-K_i}{K_+}\right)x^0\right]|{K_{+}(\epsilon')}|^{3}\, K_{+}(\epsilon')\,   \frac{\partial}{\partial P^{i}}\delta^3(\mathbf{P}-\mathbf{K})\label{C.5}
    \end{eqnarray}
Now we can see that apart from the expected bulk term there will be a non-vanishing contribution coming from the boundary, which after integrating equals to
\begin{equation}\label{C.6}
   -\frac1i\, \int_{-} \frac{d^3\mathbf{P} }{ 2\omega_{\mathbf{P}} |P_4|}
 \delta(|\mathbf{P}|^2-1) \omega_{\mathbf{P}}  |{P_{+}(\epsilon)}|^{3}  P_{i} \, a_{-\epsilon}^{\dagger}(\mathbf{P})\, a_{-\epsilon}^{\dagger *}(\mathbf{P}) +\frac1i\,  \int_{0} \frac{d^3\mathbf{P} }{ 2\omega_{\mathbf{P}} |P_4|}
 \delta(|\mathbf{P}|^2-1) \omega_{\mathbf{P}}  |{P_{+}(\epsilon)}|^{3}  P_{i} \, a_{-\epsilon}^{\dagger}(\mathbf{P})\, a_{-\epsilon}^{\dagger *}(\mathbf{P})
\end{equation}
which can be more compactly written as
\begin{equation}\label{C.7}
  \frac1i\, \int_{|\mathbf{P}|=1} \frac{d\Omega }{ 2 |P_4|}\, P_i \left( N_-(\mathbf{P}) - N_0(\mathbf{P})\right)
\end{equation}
where $d \Omega$ is the measure on the (momentum) unit sphere.

Let us now return to the bulk term. Integrating by parts, we see that when the derivative over $P_i$ hits the terms in the exponent that depend on $P$. Now we can see, after some algebra, that the only  contribution comes from the term $P_0/K_+$ term in the exponent in (\ref{C.5}) and that the contributions coming from the other terms cancel out. This contribution provides the term that cancels exactly the first term in (\ref{C.4}).  The next two terms come from the derivative hitting $P_0(\epsilon)=\omega_{\mathbf{P}}^\epsilon$ and the $a^\dag_{-\epsilon}(\mathbf{P})$. Integrating delta we finally find
\begin{equation}\label{C.8a}
  {\cal N}_{i}^{bulk} =  -\frac1i\, \sum_\epsilon\int_{\epsilon} \frac{d^3\mathbf{P} }{ 2\omega_{\mathbf{P}} |P_4|} \alpha(\epsilon) \, P_+(\epsilon)\left[ \frac{P_i(\epsilon)}{P_{0}(\epsilon)}\,N_{\mathbf{P}}(\epsilon)+
  \omega_{\mathbf{P}}\, |{P_{+}(\epsilon)}|^{3}  \, \left(\frac{\partial}{\partial P_i} \, a_{-\epsilon}^{\dagger}(\mathbf{P})\right) a_{-\epsilon}^{\dagger *}(\mathbf{P})\right]
\end{equation}
These terms are natural modifications of the ones that arise in the standard scalar field theory. Notice that again we find explicitly that the charges constructed by a rather abstract reasoning above are indeed time-independent, as they should.

This concludes our construction of conserved charges, associated with deformed Poincar\'e symmetry of our theory.

\subsection{Symplectic structure}

Let us conclude this section with a brief discussion of symplectic structure, arising from the free scalar field action on $\kappa$-Minkowski space. The symplectic potential $\rho$ is a 1-form on the space of solutions of the equation of motion and can be
easily deduced from (\ref{var}). One finds
\be
\rho=\int_{\mathbf{R}^{3}}( \Pi^{0}\delta\phi -\delta\hf^{\dagger} \Pi^{0\dagger})
\ee
The symplectic form is given by $\Omega =\delta \rho$ ($\delta $ is treated as a differential on the space of field, hence $\delta^{2}=0$)

Since on-shell we have
$$
\Pi^{0}= -\hP_{4}\hP_{0} \phi^{\dagger}
$$
we can write
\be
\Omega=- \int_{\mathbf{R}^{3}}( (\hP_{4} \hP_{0} \delta \phi^{\dagger}) \wedge \delta\phi + \delta \phi^{\dagger} \wedge (\hP_{4}^{\dagger}\hP_{0}^{\dagger}\delta\hf) )
\ee
which can be written in terms of momentum modes as follows
\bea
\Omega &=&
 \sum_{\epsilon} \int_{\epsilon} \frac{d^3\mathbf{P} }{ 2\omega_{\mathbf{P}} |P_4|}
\alpha(-\epsilon)   |{P_{+}(-\epsilon)}|^{3} \delta a_{\epsilon}^{\dagger}(\mathbf{P}) \wedge \delta a_{\epsilon}^{\dagger *}(\mathbf{P}) \\
&=&
 \sum_{\epsilon} \int_{\epsilon} \frac{d^3\mathbf{P} }{ 2\omega_{\mathbf{P}} |P_4|}
\alpha(-\epsilon)  \delta  a_{\epsilon}^{\dagger}(\mathbf{P})\wedge  \delta a_{\epsilon}^{ }(\mathbf{S^{-\epsilon}_P}) \\
&=&
 \sum_{\epsilon} \int_{\epsilon} \frac{d^3\mathbf{P} }{ 2\omega_{\mathbf{P}} |P_4|}
\alpha(-\epsilon)   \delta a_{\epsilon}^{*}(\mathbf{P}) \wedge \delta a_{\epsilon}^{}(\mathbf{P}) \\
\eea
Knowing the symplectic form, one could straightforwardly derive the Poisson bracket in the space of the Fourier field components
\be
\{a^{\dagger}_{\epsilon}(\mathbf{P}),a^{}_{\epsilon'}(\mathbf{Q})\}= \delta_{\epsilon,\epsilon'} \alpha(-\epsilon) \frac{\delta^{3}(\mathbf{S_{P}^{-\epsilon}}-\mathbf{Q})}{ |{P_{+}(-\epsilon)}|^{3}}
=\delta_{\epsilon,\epsilon'} \alpha(-\epsilon)
\left\{\begin{array}{cc}
 \delta^{3}(\mathbf{Q}\oplus\mathbf{P}),\, \mathrm{if}\,\,{\epsilon =0} \\
 \delta^{3}(\mathbf{Q}\oplus\mathbf{(-P)}),\, \mathrm{if}\,\, {\epsilon =\pm}
 \end{array}\right.
\ee
As usual this bracket will turn to commutator in the quantum theory. The quantization of our theory, and, in particular, the construction of Fock space and multiparticle states, along with the discussion of their properties will be a subject of a forthcoming paper.

\section{Discussion and conclusions}

Let us now try to summarize the results described the results we have obtained.

The first result of this paper was establishing and analyzing the connection between $\kappa$-Minkowski space, de Sitter space of momenta, and group theory. By interpreting the $\kappa$-Minkowski commutative relations as defining relation of Lie algebra of Borel group, we find that the Borel group is isomorphic to the half of de Sitter (momentum) manifold, defined by the condition $P_+>0$. In section III we saw also that the group theoretical perspective clarifies some old and well known results on differential calculus on $\kappa$-Minkowski space.

The fact that the natural momentum space for field theory is defined by the region in de Sitter space defined by $P_+>0$ has its immediate and far reaching consequences. This condition is clearly not Lorentz invariant, as it is easily seen from Figure 1, where the Lorentz orbits  in Sectors $-$ and $0$ hit the surface $P_+$ =0. Thus one could expect that any theory with such momentum space will suffer Lorentz symmetry breaking. It is also clear from Figure 1 that in the spectrum of the theory a kind of trans-Planckian particles, called Sector 0 is going to appear.

One can remedy this situation by taking the image of Sector $0$ by inversion, effectively gluing it from the bottom to Sector $-$, which changes the defining condition to $P_4>0$, which is, obviously, Lorentz invariant. (More precisely this construction amounts to assuming that the field lives on elliptic de Sitter space.) The construction of free scalar field theory with such momentum space has been presented in our recent paper \cite{LJSshort}. In that paper we found that the field theory that results from using an appropriate star product is a theory on standard Minkowski space-time that is manifestly Poincar\'e invariant and has in its spectrum a massive mode along with the tachyonic one. Fortunately, in free theory the tachyon has infinite energy, so it cannot appear in the asymptotic states. It is not completely clear yet how its presence would modify the interacting theory.

If one wants to stick strictly to the guidelines mapped out by mathematics of $\kappa$-Minkowski space, one should consider the field $\hf$ whose momentum space is Borel group, and thus a submanifold of de Sitter space defined by $P_+>0$. In this case, as it is clear from Figure 1 the breaking of Lorentz symmetry seems inevitable. Consider a set of Lorentz observers moving with respect to each other with higher and higher speed. If all of them observe a particle which appears in sector $-$ for those of the observers who find particle energy small, for the sufficiently boosted observer the particle is going to hit the boundary $P_+=0$. What the observer with even larger boost will see? She will observe no sector $-$ particle, but instead a sector $0$ one. If there is an objective way to know which sector is which (for example, in sector $-$ the particle is green and in sector $0$ it is red), this would certainly mean the breaking of Lorentz invariance (since two observers looking at the same object will not agree about its color).

Another question that we would need to address is whether all the conserved charges presented here do generalize to  the interacting case.
We would expect this to be the case but more detailed investigations are clearly needed. In particular one can wonder what happens when (using interactions) we accelerate one particle such that after some time it hits the boundary  $P_+=0$.

The conceptual interpretation of this theory is therefore not completely clear. It presumably requires deeper understanding of a role of observer and symmetries in the non-commutative setting.

\section*{Acknowledgment} For JKG and SN this research was supported in part by KBN grant 1 P03B 01828 and in part by Perimeter Institute for Theoretical Physics; for JKG this research was supported as a part of 2007-2010 research project N202 081 32/1844. For SN this work was supported by ESF-Exchange Grant-Quantum Geometry and Quantum Gravity-ref. 1420


\appendix
\section{Useful formulas}

Let us take
\begin{equation}\label{a.1}
P_{0}=\sinh{{k_0}}+\frac{\mathbf{k}^{2}}{2}\,
 e^{{k_0}}
\end{equation}
\begin{equation}\label{a.2}
 P_{i}=k_{i}\,e^{{k_0}}
\end{equation}
and define also the variable $P_4(k_0, \mathbf{k})$
\begin{equation}\label{a.3}
 P_{4}=\cosh{{k_0}} -\frac{\mathbf{k}^{2}}{2}
 e^{{k_0}}
 \end{equation}
One can easily check that
\begin{equation}\label{a.3a}
-P_0{}^2 +\mathbf{P}^{2} + P_4{}^2
=,\quad P_0+P_4=e^{{k_0}}
\end{equation}
The co-products for $P_a$ can be calculated form co-products for $k_\mu$
\begin{equation}\label{a.kPcp}
\Delta k_0 = k_0 \otimes \bbbone + \bbbone \otimes k_0, \quad \Delta k_i = k_i
\otimes \bbbone + e^{-{k_0}} \otimes k_i
\end{equation}
and read
\begin{equation}\label{a.4}
  \Delta(P_{i})=P_{i}\otimes (P_{0}+P_{4}) +\bbbone\otimes P_{i}
\end{equation}
\begin{equation}\label{a.5}
  \Delta(P_{4})=P_{4}\otimes (P_{0}+P_{4})-\sum P_{k}(P_{0}+P_{4})^{-1}\otimes P_{k}
 -(P_{0}+P_{4})^{-1}\otimes P_{0}
\end{equation}
\begin{equation}\label{a.6}
  \Delta(P_{0})=P_{0}\otimes (P_{0}+P_{4})+\sum P_{k}(P_{0}+P_{4})^{-1}\otimes P_{k}
 +(P_{0}+P_{4})^{-1}\otimes P_{0}
\end{equation}
Analogously one can derive antipodes for $P_a$  from
\be
S(k_{0})=-k_{0},\quad S(k_{i})=-k_{i}e^{k_{0}}
\ee
this reads
\begin{equation}\label{a.8}
S(P_0)=\frac{1}{P_0+P_4}-P_4= -P_{0} +\frac{\mathbf{P}^{2}}{P_{0}+P_{4}}
\end{equation}
\begin{equation}\label{a.9}
S(P_i)=\frac{-  P_i}{P_0+P_4}
\end{equation}
\begin{equation}\label{a.10}
S(P_4)=P_4
\end{equation}
The co-products and antipodes for Lorentz generators read
$$
\triangle (M_i)=M_i\otimes \bbbone+\bbbone\otimes M_i
$$
\begin{equation}\label{s31a}
    \triangle (N_i)=N_i\otimes \bbbone +e^{-{k_0}}\otimes N_i+\epsilon_{ijk}k_j\otimes M_k
\end{equation}
$$
S(M_i)=-M_i
$$
\begin{equation}\label{s18a}
 S(N_i)=-e^{{k_0}}\left(N_i-\epsilon_{ijk}k_jM_k\right).
\end{equation}
We should stress here that if we use differential representation (momenta replaced
with differentials) we have to use transposed operator:
\begin{equation}\label{s18}
S_{dif}(N_i)=-\left(N_i-\epsilon_{ijk}M_kk_j(\partial)\right)e^{{k_0(\partial)}},
\end{equation}
which of course reproduces antipode (\ref{s18a}) while acting on plane wave.


\end{document}